\documentclass[%
 reprint,
superscriptaddress,
 amsmath,amssymb,
 aps,
pra,
]{revtex4-2}

\usepackage{graphicx}
\usepackage{dcolumn}
\usepackage{bm}
\usepackage{physics}
\usepackage[english]{babel}

\usepackage{tcolorbox}
\usepackage{orcidlink}

\newcommand{\dnorm}[1]{\abs{\abs{#1}}_\diamond}

\tcbuselibrary{theorems}

\newtcbtheorem[]{mytheo}{}%
{colback=blue!5,colframe=blue!50!black,fonttitle=\bfseries}{th}

\begin{document}

\title{Bounding the systematic error in quantum error mitigation due to model violation}

\author{L.~C.~G.~Govia}
\affiliation{IBM Quantum, IBM Almaden Research Center, San Jose, CA 95120, USA}
\author{S.~Majumder}
\affiliation{IBM Quantum, IBM T.~J.~Watson Research Center, Yorktown Heights, NY 10598, USA}
\author{S.~V.~Barron}
\affiliation{IBM Quantum, IBM T.~J.~Watson Research Center, Yorktown Heights, NY 10598, USA}
\author{B.~Mitchell}
\affiliation{IBM Quantum, IBM Almaden Research Center, San Jose, CA 95120, USA}
\author{A.~Seif}
\affiliation{IBM Quantum, IBM T.~J.~Watson Research Center, Yorktown Heights, NY 10598, USA}
\author{Y.~Kim}
\affiliation{IBM Quantum, IBM T.~J.~Watson Research Center, Yorktown Heights, NY 10598, USA}
\author{C.~J.~Wood}
\affiliation{IBM Quantum, IBM T.~J.~Watson Research Center, Yorktown Heights, NY 10598, USA}
\author{E.~J.~Pritchett}
\affiliation{IBM Quantum, IBM T.~J.~Watson Research Center, Yorktown Heights, NY 10598, USA}
\author{S.~T.~Merkel}
\affiliation{IBM Quantum, IBM T.~J.~Watson Research Center, Yorktown Heights, NY 10598, USA}
\author{D.~C.~McKay}
\affiliation{IBM Quantum, IBM T.~J.~Watson Research Center, Yorktown Heights, NY 10598, USA}



\begin{abstract}
Quantum error mitigation is a promising route to achieving quantum utility, and potentially quantum advantage in the near-term. Many state-of-the-art error mitigation schemes use knowledge of the errors in the quantum processor, which opens the question to what extent inaccuracy in the error model impacts the performance of error mitigation. In this work, we develop a methodology to efficiently compute upper bounds on the impact of error-model inaccuracy in error mitigation. Our protocols require no additional experiments, and instead rely on comparisons between the error model and the error-learning data from which the model is generated. We demonstrate the efficacy of our methodology by deploying it on an IBM Quantum superconducting qubit quantum processor, and through numerical simulation of standard error models. We show that our estimated upper bounds are typically close to the worst observed performance of error mitigation on random circuits. Our methodology can also be understood as an operationally meaningful metric to assess the quality of error models, and we further extend our methodology to allow for comparison between error models. Finally, contrary to what one might expect we show that observable error in noisy layered circuits of sufficient depth is not always maximized by a Clifford circuit, which may be of independent interest.

\end{abstract}

\maketitle

\section{Introduction}

Any potential advantage from quantum computers will ultimately depend on the quality, speed, and scale at which they are able to run computations. As scale increases across a variety of hardware platforms \cite{utility,Google23,Bluvstein24,Balewski24,RyanAnderson24} a prominent focus in the near-term has been on improving the quality of operations at large system sizes that cannot yet implement fault-tolerant quantum error correction. Error mitigation is a catch-all term that refers to a loosely connected set of protocols that use additional quantum circuits, classical pre/post-processing, or both to improve the accuracy of a target computation \cite{Error_mit_rmp}, thereby increasing quality at the cost of reduced speed. This has enabled demonstrations at a scale and quality previously unattainable \cite{utility,Shtanko23,Hoke23,Yu23,Chen23-ibm,Farrell24_1,Farrell24_2}.

Several promising error mitigation schemes rely on characterization of the error channel associated with the layers of a target circuit, for example Refs.~\cite{Temme17,Endo18,Strikis21,Ferracin22,Berg2023,Guo22,Filippov23}. As such, the accuracy of the error model obtained through characterization directly determines the quality of the mitigated computation result. However, since the true error channel of the circuit is unknown, quantifying the accuracy of the characterized error model is a challenging task. Moreover, ideally one would want to measure model-accuracy at no further experimental cost, both to avoid a further reduction in computation speed, and to minimize the impact of fluctuations or drift of the error model on the analysis \cite{kim2024}.

In this work, we study the impact of inaccurate error models on error mitigation protocols. To do so, we use the concept of model violation, which captures how well an error model is able to reproduce experimental data, and has previously been used to assess the quality of quantum gate-set tomography \cite{wildcard}. We derive upper bounds to the inaccuracy of error mitigation due to model violation by quantifying how incorrect model parameters propagate into error mitigated expectation values. While our primary focus is answering the question of how well an error channel can be mitigated by its model, the methodology of model violation can also be used to understand how one model mitigates another. This has application in, for example, quantifying the impact on error mitigation of temporal fluctuations or drift in the error channels of a device \cite{kim2024}. While previous work has focused on estimating the impact of model violation on error mitigation of ``typical'' circuits \cite{Filippov24}, here we focus on worst-case upper bounds. 

The manuscript is organized as follows. In section \ref{sec:ecmv} we review error-model based error mitigation and the sources of model violation, and in section \ref{sec:mv_bounds} we discuss the requirements for our upper bounds. In section \ref{sec:PEC} we present a rigorous and in section \ref{sec:Clif} a heuristic upper bound for model violation in probabilistic error cancellation \cite{Berg2023}. We deploy these upper bounds in simulation in section \ref{sec:numsims} and on an IBM Quantum processor in section \ref{sec:exp}. We present our method for model comparison in section \ref{sec:stability}, and make concluding remarks in section \ref{sec:conc}. The six appendices contain supporting details and information, and we would like to highlight Appendices \ref{sec:PEA} and \ref{sec:TEM} that extend our methods to other error-model based mitigation protocols, namely probabilistic error amplification \cite{utility}, and tensor-network error mitigation \cite{Filippov23}.

\section{Error mitigation and model violation}
\label{sec:ecmv}

Let $\mathcal{C}$ denote the ideal target circuit of interest and $\tilde{\mathcal{C}}$ its error-prone implementation on hardware, which we will often call the ``noisy'' circuit. We assume that $\mathcal{C}$ has a layer structure, consisting of a finite number of error-prone ``hard'' layers (these often contain two-qubit gates) interleaved by error-free ``easy'' layers (often single-qubit gates). Such a circuit structure is common in quantum simulation \cite{Georgescu14}, e.g.~by Trotterization, quantum error correction \cite{Terhal15}, many variational algorithms \cite{Cerezo21}, and any circuit employing Pauli twirling or randomized compiling \cite{Wallman16}. Furthermore, we assume that the error channel for each hard-layer is a Pauli-stochastic channel, 
\begin{align}
    \mathcal{E}(\rho) = \sum_{j=0}^{4^n-1}\nu_jP_j \rho P_j, \label{eqn:pc}
\end{align}
where $P_j$ are the $n$-qubit Pauli operators. This can be enforced by Pauli twirling the hard-layers, i.e.~inserting random Pauli gates before and after the hard-layers \cite{Berg2023,Bennett96,Knill04,Kern05,Geller13,Wallman16}.

For an initial state $\rho$ and observable $O$, we consider error mitigation protocols $\mathbb{M}$ for which the goal is to return an estimate $\tilde{o} = \mathbb{M}\left(\tilde{\mathcal{C}},\rho,O,\{\hat{\mathcal{E}}_k\}\right)$ to the ideal expectation value $o = {\rm Tr}\left[\mathcal{C}(\rho)O\right]$ that is as close as possible in absolute value, i.e.~minimizes $\delta_o = \abs{o-\tilde{o}}$. We focus on error mitigation protocols that use knowledge of the hard-layer error channels, described by the set $\{\hat{\mathcal{E}}_k\}$ of models for the error channels $\{\mathcal{E}_k\}$ of each hard-layer indexed by $k$. Examples include probabilistic error cancellation (PEC) \cite{Berg2023}, probabilistic error amplification (PEA) \cite{utility}, and tensor-network error mitigation (TEM) \cite{Filippov23}.

The inaccuracy $\delta_o$ will contain both a statistical component, and a systematic error or bias. The statistical component results from the fact that the error mitigation protocols we consider increase accuracy at the cost of increased variance (for a fixed number of samples) in their estimate. The increase in samples required to overcome this increased variance, commonly referred to as the overhead of the error mitigation protocol, has been well studied in previous work \cite{Endo18,Regula2021,Takagi22,Berg2023,Filippov24}. The systematic error, i.e.~the inaccuracy that remains in the infinite sample limit, can have many sources. It is often intrinsic to method of the protocol, but even protocols such as PEC that are bias-free when perfectly implemented are susceptible to bias due to inaccurate error models.

The above mentioned protocols rely on a learning phase that produces a model for the error channel of each hard-layer. Pauli-learning \cite{Berg2023,Kimmel14,Helsen19,Erhard19,Harper20} is a commonly deployed technique to produce these models, but in practice it is susceptible to several sources of model violation:
\begin{enumerate}
    \item \emph{The error channel is not Pauli-stochastic}. This would occur if an insufficient number of randomizations were used, or if the strategy for choosing random Pauli gates is biased.
    
    \item \emph{Non-Markovian error}. Error that is  correlated in time, or depends on the context (e.g.~state) of systems not included in the error model violates the assumption that the error channel is Markovian and can be described by a single completely-positive and trace preserving (CPTP) channel. Examples include fluctuations or drift of error rates with time. Such error sources can induce model violation that is often a function of circuit depth. 

    \item \emph{Out-of-model error}. Pauli-learning has exponential complexity unless further assumptions are made that constrain the interaction map between qubits. For example, restricting interactions to be pairwise between intentionally coupled qubits \cite{Berg2023}. Error processes not respecting these constraints would result in model violation.

    \item \emph{Learning degeneracy}. Using the error amplification sequences of Pauli-learning, it is impossible to measure all Pauli fidelities uniquely in a way that separates system preparation and measurement error (SPAM) from layer error \cite{Berg2023,Chen23}. Instead, some Pauli fidelities can only be measured as paired products. Breaking this degeneracy requires model assumptions that if incorrect would not induce model violation in the Pauli-learning circuits, but could when applied to other circuits.

    \item \emph{Statistical uncertainty}. Due to finite sampling of experiments in Pauli-learning, there will be inherent uncertainty in the values of the model coefficients. The model violation due to this should decrease as more samples are taken during Pauli-learning, but will play a role in any realistic experiment, especially when learning time is constrained by other issues such as the timescale of error channel drift.

\end{enumerate}
Items 1 and 2 listed above are sources of model violation that are properties of the error channel of the system, while items 3-5 are a result of the learning procedure. In the following, our goal is to obtain upper bounds for the portion of $\delta_o$ (the error in our mitigated expectation value) that is due to model violation.

\section{Model violation error bounds}
\label{sec:mv_bounds}

Estimations of model violation of hard-layers can be swamped by the often significant error in initial state preparation, or in measurement (collectively called SPAM). In practice one often treats SPAM with a separate error mitigation procedure in post-processing, e.g.~Ref.~\cite{Berg22}, and systematic error due to incorrect SPAM estimation should be analyzed separately. As such, to measure model violation of the error model for the hard-layers we need data free from SPAM. 

For this, we consider the Pauli fidelities measured during the first stage of a Pauli-learning scheme, which we denote $f^{\rm meas}_P$ for Pauli $P$. The Pauli fidelities fully characterize a Pauli-stochastic channel, and are related to the $\nu_j$ of Eq.~\eqref{eqn:pc} by a Fourier transform \cite{Chen23}. They are commonly extracted from exponential fits as a function of depth to the decaying signal obtained from the procedure: i) prepare a Pauli eigenstate, ii) repeat a specific hard-layer $d$ times, interleaving random single-qubit Pauli gates between hard-layers to implement a Pauli twirl, and iii) measure in the corresponding Pauli basis. Using $f^{\rm meas}_P$ as our ``ground-truth'' will miss sources of model violation that cause the same deviation from exponential decay in all decay curves, but we consider such situations pathological.

Note also that to account for the degeneracy inherent to any Pauli-learning scheme, i.e.~where for some Pauli operators only the product  $f^{\rm meas}_{P_1}f^{\rm meas}_{P_2}$ can be learnt in a SPAM-free manner \cite{Chen23}, we simply set both Pauli fidelities to be $\sqrt{f^{\rm meas}_{P_1}f^{\rm meas}_{P_2}}$. This choice likely itself introduces model violation, but we will not explore alternate options in this work. Alternatively, one can isolate model violation robust to this assumption by studying only test circuits that are invariant under how the degeneracy is broken, such as even depth Pauli-learning circuits.

To estimate model violation, we compare the measured Pauli fidelities to what the model predicts the Pauli fidelities should be, which we denote $f^{\rm mod}_P$.
Although we learn by fitting the measured Pauli fidelities, $f^{\rm meas}_P$, they are not themselves an error model. The next step of any learning protocol is to fit an error model with free parameters $\vec{\lambda}$ to the measured Pauli fidelities, which we do by minimizing an appropriate cost function that compares $f^{\rm meas}_P$ to $f^{\rm mod}_P = F(\vec{\lambda},P)$. For example, in the Pauli-Lindblad model we consider here \cite{Berg2023}, $\vec{\lambda}$ are the coefficients in an error generator such that the error channel is given by
\begin{align}
    \mathcal{E} = e^\mathcal{L},~\mathcal{L}(\rho) = \sum_j \lambda_j\left(P_j\rho P_j - \rho\right). \label{eqn:PLmodel}
\end{align}
From this we obtain that $f^{\rm mod}_P = \exp(-2\vec{M}_{P}\cdot\vec{\lambda})$, where the vector $\vec{M}_P$ has elements $[\vec{M}_{P}]_j = 1$ if Pauli $P$ and $P_j$ anticommute, and zero otherwise. We obtain $\vec{\lambda}$ by a non-negative least squares optimization of $\abs{f^{\rm meas}_P - f^{\rm mod}_P}^2$.

As the fitting procedure is constrained with fewer or exactly as many parameters $\vec{\lambda}$ as measured $f^{\rm meas}_P$, the fidelities predicted by the model $f^{\rm mod}_P$ may not match the fidelities from the data $f^{\rm meas}_P$. This is precisely the signature we exploit to quantify model violation. In the following, we describe our procedure for estimating the impact of model violation on the systematic error component of $\delta_o$, using only comparisons between the measured and model Pauli fidelities. We focus on probabilistic error cancellation, and consider the extension of our techniques to probabilistic error amplification in Appendix \ref{sec:PEA} and tensor-network error mitigation in Appendix \ref{sec:TEM}.

\section{Rigorous Diamond Distance Model Violation Bounds for PEC}
\label{sec:PEC}

By inserting correction gates sampled randomly from the distribution defined by the error model learnt for a hard-layer, probabilistic error cancellation inverts (on average) the error channel of the hard-layer. If PEC works perfectly the average of all mitigated circuits is equivalent to the ideal circuit. Typically, the learnt error model will not exactly match the error of the hard-layer, and our goal is to quantify the impact this has on PEC using model violation. 

From here on we assume the infinite sample limit for mitigation circuits, such that the inaccuracy $\delta_o$ has only a systematic component. For a circuit with a single hard-layer $\mathcal{U}$, we can write $\delta_o$ for PEC as
\begin{align}
    \delta_o &= \abs{{\rm Tr}\left[\mathcal{U}(\rho)O\right] - {\rm Tr}\left[\tilde{\mathcal{U}}\hat{\mathcal{E}}^{-1}(\rho)O\right]} 
\end{align}
where $\hat{\mathcal{E}}$ is the learnt error model. From here on we use $\Lambda\Gamma(\rho)$ as shorthand for the composition $\Lambda(\Gamma(\rho)) = \Lambda\circ\Gamma(\rho)$. Note that $\tilde{\mathcal{U}}\hat{\mathcal{E}}^{-1}(\rho)$ is still Hermitian because $\hat{\mathcal{E}}^{-1}$ is Hermiticity preserving for Pauli-stochastic $\hat{\mathcal{E}}$. 

Since ${\rm Tr}\left[A^\dagger B\right] \leq \abs{\abs{A}}_1 \abs{\abs{B}}_\infty$, and $\abs{\abs{O}}_\infty = 1$ for Pauli observables, we can bound $\delta_o$ from above by the trace distance between output states
\begin{align}
    \delta_o \leq \abs{\abs{\mathcal{U}(\rho) - \tilde{\mathcal{U}}\hat{\mathcal{E}}^{-1}(\rho)}}_1 \leq \dnorm{\mathcal{U} - \tilde{\mathcal{U}}\hat{\mathcal{E}}^{-1}},
\end{align}
which we can bound (by definition) from above by the diamond distance between maps \cite{Watrous_2018} 
\begin{align}
    \dnorm{\Lambda} = \max\left\{\norm{\left(\Lambda\otimes\mathcal{I}\right)(X)}_1 : X \in L(\mathcal{H}^{\otimes 2}), \norm{X}_1 \leq 1 \right\}.
\end{align}
Since $\tilde{\mathcal{U}}\hat{\mathcal{E}}^{-1}$ is not completely-positive (CP) we use the more general definition of the diamond norm that optimizes over unit-bounded trace-norm matrices instead of density matrices. Finally, writing $\tilde{\mathcal{U}} = \mathcal{U}\mathcal{E}$ we can use sub-multiplicativity of the diamond norm to arrive at
\begin{align}
    \delta_o \leq \dnorm{\mathcal{I} - \mathcal{E}\hat{\mathcal{E}}^{-1}},
    \label{eqn:fundnorm}
\end{align}
which bounds the systematic error due to model violation by a quantity that depends only on the learnt $(\hat{\mathcal{E}})$ and actual $(\mathcal{E})$ error models, which is independent of choice of observable and initial state.

For a multi-layer circuit we can similarly derive the more complicated expression (see Appendix \ref{app:DNbounds})
\begin{align}
    \delta_o \leq \sum^l_j \dnorm{\mathcal{I} - \mathcal{E}_j\hat{\mathcal{E}}_j^{-1}}\prod_k^{j-1}\dnorm{\mathcal{E}_k\hat{\mathcal{E}}_k^{-1}}, \label{eqn:PECgen}
\end{align}
where the product of diamond norms of the $\mathcal{E}_k\hat{\mathcal{E}}_k^{-1}$ maps appears because they are not CP. Both $\mathcal{E}_k$ and $\hat{\mathcal{E}}_k^{-1}$ have diagonal Pauli transfer matrices, the former since it is assumed to be a Pauli-stochastic channel and the latter because it is the inverse map of a Pauli-stochastic channel. Therefore, defining $\mathcal{V}_k = \mathcal{E}_k\hat{\mathcal{E}}_k^{-1}$, we see that it will also have the structure of a Pauli-stochastic channel,
\begin{align}
    \mathcal{V}(\rho) = \sum_j \nu_j P_j\rho P_j,
\end{align}
but with $\nu_j$ not necessarily positive or normalized (for notational convenience we have dropped the subscript $k$ when it is not needed to distinguish layers). We dub such channels \emph{Pauli Stochastish}. Pauli Stochastish channels have diagonal Pauli transfer matrices, and the ones we consider in this work are trace preserving, such that the first diagonal element $r_0 = 1$. The rest will be the ratio between the Pauli fidelity of the actual noise channel and the model, $r_k = f^{\rm meas}_{P_k}/f^{\rm mod}_{P_k}$.

It is instructive to consider the intuition behind the form of Eq.~\eqref{eqn:PECgen}. It consists of a sum over diamond distances of each layer, $\dnorm{\mathcal{I} - \mathcal{V}_j}$, with a weighting factor formed by a product of diamond norms from previous layers $\dnorm{\mathcal{V}_k}$. The diamond distance measures how close the mitigated channel at each layer is to identity, and thus the sum of such distances over layers is naturally a measure of the systematic error. The weighting factor comes about because the mitigated map may not be CP, which in practice is a statement that PEC can over-correct errors as well as under-correct them. The dependence on previous layers points out that PEC can over-correct error of earlier layers to compensate for under-correction at the current layer. Thus, for a given circuit, failed mitigation at each layer can conspire to result in an overall lower systematic error in the full circuit than for any one layer. This is an example of a ``horoscope effect'' \footnote{We thank our colleague David Layden for drawing this analogy.}, where PEC can return the right answer for the wrong reasons. It is impossible for this to happen for every circuit, and the upper bound of Eq.~\eqref{eqn:PECgen} shows that the worst case is when this over/under-correction conspires to amplify the systematic error. Thus, while some circuits may be deceivingly accurate, generically good performance still requires accurate mitigation layer-by-layer.

\subsection{Upper Bounding Systematic Error}

We now aim to bound the terms in Eq.~\eqref{eqn:PECgen} by quantities that can be estimated from experimental data, and in particular data already available from Pauli-learning. The procedure is outlined below, with details in the rest of this section and Appendix \ref{app:bounds}.

\begin{mytheo*}{Procedure to Estimate Systematic Error Upper-Bound}
\begin{enumerate}
    \item Obtain measured Pauli fidelities from Pauli-learning exponential fits: $f_P^{\rm meas}$.
    \item Fit data $f_P^{\rm meas}$ to a model, e.g.~Eq.~\eqref{eqn:PLmodel}, to obtain model coefficients $\vec{\lambda}$.
    \item Compute model-predicted Pauli fidelities: $f_P^{\rm mod}$. $f_P^{\rm mod} \neq f_P^{\rm meas}$ in the presence of model violation.
    \item Estimate the systematic error upper bound of Eq.~\eqref{eqn:PECgen} using Eq.~\eqref{eqn:pecbound_dne} or Eq.~\eqref{eqn:pecbound_2norm}.
\end{enumerate}
\end{mytheo*}

Using a result from Ref.~\cite{Watrous13} that relates the diamond norm to the maximum output fidelity, we can express the diamond norm of $\mathcal{V}$ as (see Appendix \ref{app:dnormPS} for further details)
\begin{align}
    \dnorm{\mathcal{V}} = \sum_{j=0} \abs{\nu_j}.
\end{align}
Unfortunately, we do not have access to $\nu_j$. Instead, we upper bound $\dnorm{\mathcal{V}}$ by using the fact that the diamond norm is sub-multiplicative to obtain (see Appendix \ref{app:invdnorm})
\begin{align}
    \dnorm{\mathcal{V}} = \dnorm{\mathcal{E}\hat{\mathcal{E}}^{-1}} \leq \dnorm{\mathcal{E}}\dnorm{\hat{\mathcal{E}}^{-1}} \leq \gamma, \label{eqn:Vdnorm}
\end{align}
where $\gamma$, as defined in Ref.~\cite{Berg2023}, is a measure of the model channel and can be calculated directly from the model coefficients via $\gamma = \exp(2\sum_i \lambda_i)$. 

Upper bounding $\dnorm{\mathcal{I} - \mathcal{V}}$ is more complicated. It can be expressed as
\begin{align}
    \dnorm{\mathcal{I} - \mathcal{V}} = \abs{1-\nu_0} + \sum_{m=1} \abs{\nu_m},
\end{align}
but as mentioned previously we do not have access to $\nu_j$. We only have access to a subset of the Pauli fidelity ratios $r_k$, which are related to $\nu_j$ via the Fourier transform
\begin{align}
    \nu_m = \frac{1}{4^n}\sum_k \left(-1\right)^{\left<m,k\right>}r_k, \label{eqn:FTnu}
\end{align}
where $\left<m,k\right> =0$ if Paulis $P_m$ and $P_k$ commute and one if they anticommute. We now present two approaches to upper bound $\dnorm{\mathcal{I} - \mathcal{V}}$. 

The first upper bounds $\sum_{j=1} \abs{\nu_m}$ by a function of $\gamma$ to obtain the overall upper bound $\delta_o \leq \delta_\gamma$
\begin{align}
    \delta_o \leq \sum_j \left( \abs{1-\nu_0^{(j)}} + \gamma^{(j)} - \nu_0^{(j)}\right)\prod^{j-1}_k \gamma^{(k)} \equiv \delta_\gamma,
    \label{eqn:pecbound_dne}
\end{align}
where the superscript $(j)$ indicates the $j$'th layer of the circuit. $\nu^{(j)}_0 = (\sum_k r^{(j)}_k)/4^n$ would be the process fidelity if $\mathcal{V}_j$ was CPTP. As a result of the fact that we replace $\dnorm{\mathcal{V}}$ by $\gamma$, the expression in Eq.~\eqref{eqn:pecbound_dne} does not go to zero even for $\mathcal{V} = \mathcal{I}$. Thus, this upper bound is usually more useful in the large model violation regime, and typically becomes looser as the quality of error mitigation improves. Details of the derivation of Eq.~\eqref{eqn:pecbound_dne} can be found in Appendix \ref{subsec:dnpf}.

The second upper bounds $\sum_{m=1} \abs{\nu_m}$ by the two-norm of $\nu_m$ expressed in terms of $\nu_0$ and the Pauli-fidelity ratios $r_k = f^{\rm meas}_{P_k}/f^{\rm mod}_{P_k}$ as
\begin{align}
    \delta_o \leq \sum_j \left( \abs{1-\nu_0^{(j)}} + T(\{r^{(j)}_k\})\right)\prod^{j-1}_k \gamma^{(k)} \equiv \delta_2,
    \label{eqn:pecbound_2norm}
\end{align}
where
\begin{align}
    \nonumber T(\{r^{(j)}_k\}) = \frac{4^n-1}{4^n}\sqrt{\sum_{k}r_k\left(r_k - \frac{1}{4^n-1}\sum_{m \neq k}r_m\right)}.
\end{align}
The term $T(\{r^{(j)}_k\})$ can be interpreted as a measure of how much any one Pauli fidelity ratio deviates from an average-like quantity. In contrast to $\delta_\gamma$, $\delta_2$ does indeed go to zero for perfect error mitigation, where $\mathcal{V}$ is the identity channel and $r_k=1~\forall~P_k$. Details of the derivation of Eq.~\eqref{eqn:pecbound_2norm} can be found in Appendix \ref{subsec:2norm}.

What both approaches have in common is that they can be computed from Pauli-learning data at no further experimental cost. To guarantee a strict upper bound we would need to collect all $4^n$ Pauli fidelity ratios, $r_k = f^{\rm meas}_{P_k}/f^{\rm mod}_{P_k}$, which is inefficient in sampling cost. As such, any upper bound we report should be considered an approximate estimate to the true upper bound. Of course, if further Clifford circuits are sampled then the estimate could be improved.

The mitigation map $\mathcal{V}$ is not guaranteed to be CP, but if it is then $\dnorm{\mathcal{V}} = 1$, and $\dnorm{\mathcal{I} - \mathcal{V}} = 2(1-\nu_0)$ \cite{Magesan12}, where $\nu_0 = (\sum_j r_j)/4^n$ is the process fidelity of the channel $\mathcal{V}$. In this case, Eq.~\eqref{eqn:fundnorm} simplifies to
\begin{align}
    \delta_o \leq 2\sum_j^l \left(1-\nu_0^{(j)}\right)~~\left[{\rm CPTP}\right],
\end{align}
where the process fidelity of each layer can be efficiently estimated via randomized benchmarking style experiments \cite{layer_fid,Erhard19}. Unfortunately, we have no way of guaranteeing that the mitigated map is CPTP. It is also worth keeping in mind that any upper bound for $\dnorm{\mathcal{I} - \mathcal{V}}$ can also be used to upper bound $\dnorm{\mathcal{V}}$, since
\begin{align}
    \dnorm{\mathcal{V}} = \dnorm{\mathcal{I} - \left(\mathcal{I} - \mathcal{V}\right)} \leq 1 + \dnorm{\mathcal{I} - \mathcal{V}}. \label{eqn:Vdnorm_2}
\end{align}
However, as we will only be able to produce an estimate to the upper bounds for $\dnorm{\mathcal{I} - \mathcal{V}}$ from available experimental data, choosing $\dnorm{\mathcal{V}} \leq \gamma$ is more rigorous. 

\section{Heuristic Clifford Circuit Model Violation Bounds for PEC}
\label{sec:Clif}

When $r_k = r~\forall~P_k\neq\mathbb{I}$ it is straightforward to show the $\delta_2$ upper bound has a very simple form is given by 
\begin{align}
    \nonumber\dnorm{\mathcal{I} - \mathcal{V}} &\leq \abs{1-\nu_0} + \frac{4^n-1}{4^n}\abs{1-r} = 2\frac{4^n-1}{4^n}\abs{1-r}.
\end{align}
Such a mitigated map is an example of a global depolarizing map: a Pauli Stochastish channel with Pauli fidelities $r_{\mathbb{I}} = 1$, and $r_{P_k} = r~\forall~P_k\neq \mathbb{I}$, with $r\in\mathbb{R}$. Direct calculation of the diamond distance to the identity for such a channel gives
\begin{align}
    \dnorm{\mathcal{I} - \mathcal{D}_r} = 2\frac{4^n-1}{4^n}\abs{1-r} = 2\abs{1-\nu_0}, \label{eqn:dd_dep}
\end{align}
such that the upper bound is saturated. Of course, if the mitigated map $\mathcal{V}$ was CPTP its diamond distance to identify would also be that of Eq.~\eqref{eqn:dd_dep} with $\nu_0$ the process fidelity. Systematic error greater than Eq.~\eqref{eqn:dd_dep} is thus an indication of the non-physicality of the mitigated map. 

However, for a global depolarizing map a trace-less observable (e.g.~a Pauli operator) cannot have error larger than $\abs{1-r}$, and our observables of interest cannot saturate the diamond distance upper bound. In light of this, we define a heuristic upper bound by looking at the product of the largest (smallest) Pauli fidelity ratio at each layer. We define as a heuristic upper bound for $\delta_o$ by whichever of these two products is furthest from one, i.e.
\begin{align}
    \delta_o \lessapprox \max\left[\abs{1-\prod_k\max_j r_j^{(k)}},\abs{1-\prod_k\min_j r_j^{(k)}}\right],
    \label{eqn:pecbound_wce}
\end{align}
where the product $k$ is over layers and the optimizations $j$ over Pauli fidelity ratios at each layer. 

We call this the worst-case Clifford circuit bound, as both quantities in Eq.~\eqref{eqn:pecbound_wce} can be obtained with circuits formed entirely from Clifford gates. Since Clifford gates transform Pauli operators to other Pauli operators, starting with a Pauli operator $O$, the mitigated version of a Clifford circuit simply picks up the Pauli fidelity ratio at each layer corresponding the the transformed version of $O$ appearing at that layer. Thus, by correctly choosing the Clifford gates at each layer, we can ensure that we obtain either of the two products of Pauli fidelity ratios in Eq.~\eqref{eqn:pecbound_wce}.

\begin{figure}[t!]
    \centering
    \includegraphics[width=\columnwidth]{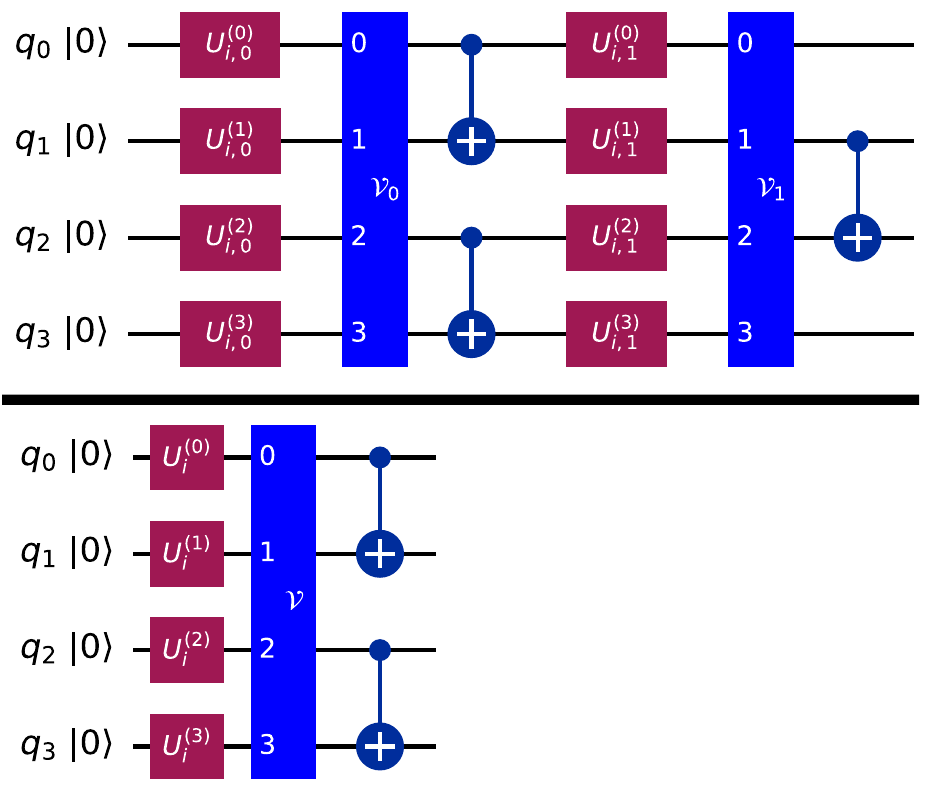}
    \caption{{\bf (Upper panel)} Building block of the 4-qubit circuits used for the model parameter uncertainty simulations of Section \ref{sec:uncertainty}. The mitigated channels $\mathcal{V}_0$ and $\mathcal{V}_1$ are built from an error and model channel that both respect the sparse Pauli-Lindblad approximation with error generators respecting the device topology. {\bf (Lower panel)} Building block of the 4-qubit circuits used for the out-of-model error simulations of Section \ref{sec:xtalk}. The mitigated channel $\mathcal{V}$ is built from an actual error channel consisting of long-range $ZIIZ$ cross-talk added to a random sparse Pauli-Lindblad error channel, and a model obtained by simulating the Pauli-learning procedure on the actual error channel.}
    \label{fig:circs}
\end{figure}

\begin{figure*}[t!]
    \centering
    \includegraphics[width=2\columnwidth]{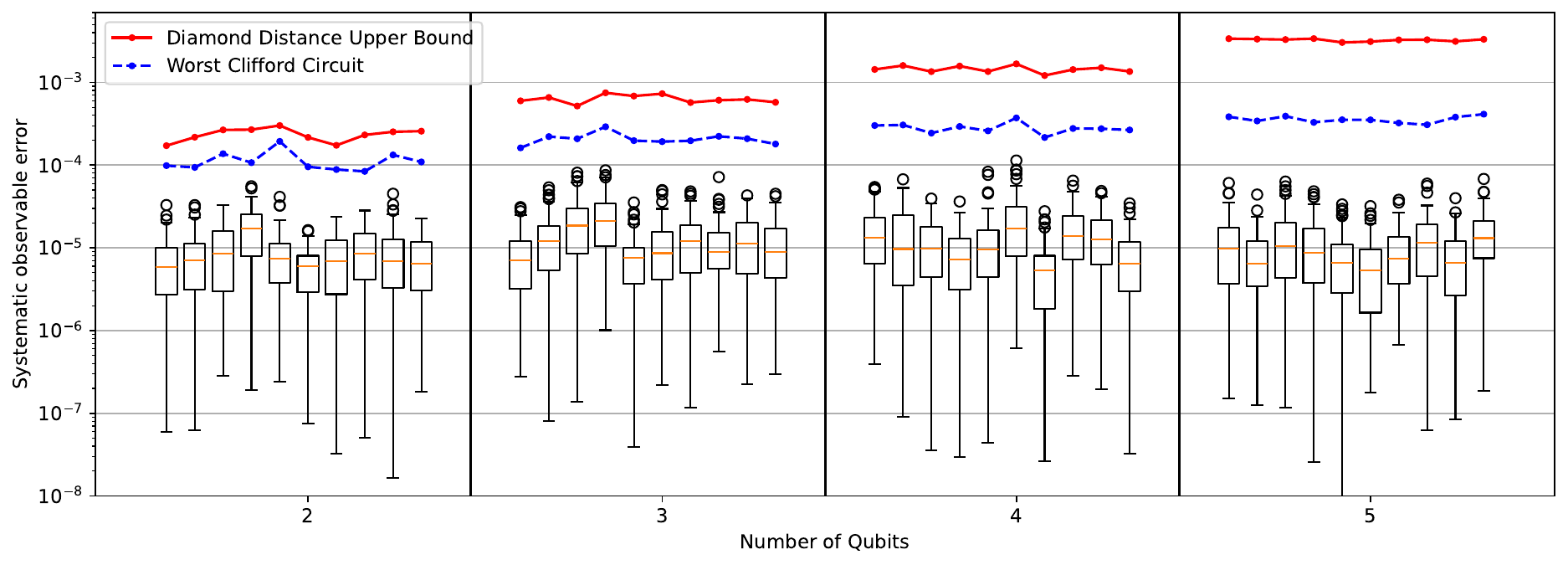}
    \caption{{\bf Learning uncertainty simulations.} 
    Box and whisker plots of the distribution of systematic error after mitigation, $\delta_o$, for random perturbations to a fixed error model, for $2$, $3$, $4$, and $5$ qubits. The fixed error model used for mitigation has model parameters $\lambda_k$ drawn uniformly at random from $\left[10^{-5},10^{-4}\right]$, which is chosen to match current expectations for experimental gate fidelities. For each qubit-number, we consider 10 perturbed error models, each with model parameters $\{\lambda_k + \epsilon_k\}$ for $\epsilon_k$ drawn uniformly at random from $\left[10^{-6},10^{-5}\right]$. 
    The circuit structure is that shown in Fig.~\ref{fig:circs} (upper panel) repeated to a depth of 20. For the fixed initial state $\ket{0}^{\otimes n_q}$, we sample 100 random circuits and for each measure a random $n_q$-qubit Pauli observable. The red solid line plots the diamond distance upper bound to the systematic error of any circuit, the smaller of $\delta_\gamma$ of Eq.~\eqref{eqn:pecbound_dne} or $\delta_2$ of Eq.~\eqref{eqn:pecbound_2norm}. The blue dashed line plots the worst-case Clifford circuit systematic error of Eq.~\eqref{eqn:pecbound_wce}. Lower quartiles of the box and whisker plots are cut from view to focus on the more important information at higher systematic error.}
    \label{fig:uncertianty}
\end{figure*}

We emphasize that even if estimated perfectly, this worst-case Clifford circuit is not a true upper bound, and one can construct circuits and error models with $\delta_o$ larger than the worst-case Clifford circuit (see Appendix \ref{app:Clif_WC} for further details). However, the numerical simulations of Section \ref{sec:numsims} seem to indicate that such situations are rare, and as we will also discuss for some physically motivated error models it can be shown that Clifford circuits exactly or nearly saturate the diamond distance upper bounds. As such, we consider the worst-case Clifford circuit to be a useful heuristic for mitigation performance, especially since the expectation value of the ideal or mitigated Clifford circuit (with Pauli stochastic error) can be evaluated efficiently with classical simulation.

\section{Numerical Simulations of Model Violation}
\label{sec:numsims}

\subsection{Model parameter uncertainty}
\label{sec:uncertainty}

The first example of model violation we consider is the situation where the actual error channel satisfies all of our assumptions (a Markovian Pauli-stochastic channel with only pairwise generator terms respecting the device topology), but the learnt model parameters differ from the actual error channel generator rates due to uncertainty in the learning procedure. To simulate this, we fix a set of model coefficients $\{\lambda_k\}$ to act as our learnt model, and then consider perturbed sets $\{\lambda_k + \epsilon_k\}$ to serve as the actual error channels, with $\epsilon_k$ drawn uniformly at random and  $\abs{\epsilon_k} < \lambda_k$. For each of the actual error channels (formed by the perturbed sets $\{\lambda_k + \epsilon_k\}$) we then simulate the act of mitigating this error with the learnt model $\{\lambda_k\}$, by defining a mitigated map $\mathcal{V}$ formed by the composition of the perturbed error channel and the learnt model inverse. For our simulations, we use a circuit structure consisting of random single-qubit gates interleaving the even/odd {\tt CNOT} layers of a line of qubits. We consider distinct mitigated maps for the even and odd layers, which act before the layer. An example of a depth one instance of this circuit structure on four qubits is shown in the upper panel of Fig.~\ref{fig:circs}.

Figure \ref{fig:uncertianty} shows the results of our simulations for up to 5 qubits. For each qubit number, we randomly sample 10 sets of perturbed model coefficients to act as the actual error channels, and consider mitigating each of these with the unperturbed model coefficients functioning as the learnt error model. We plot the distribution of systematic error over random circuits for each of these 10 simulated mitigation experiments per qubit-number. As expected, the rigorous diamond distance upper bound exceeds the systematic error for any simulated random (generically non-Clifford) circuit, even the high fliers indicated by open circles. Even though this is not provably true in general, for these simulations the worst-case Clifford circuit is also a good upper bound, and the gap between this and the diamond distance upper bound appears to be increasing with qubit number. As we expect non-Clifford circuits with systematic error within that gap to be rare and contrived, the widening of this gap speaks to the importance of the worst-case Clifford circuit as a heuristic upper bound that remains useful for larger system sizes and circuit depths.

\subsection{Long-range cross-talk}
\label{sec:xtalk}

The second example of model violation that we consider is an out-of-model error. We consider a connected line of 4 qubits and an actual error model that is pairwise local (respecting the connectivity of the line) except for the addition of a long-range cross-talk error between qubits $q_0$ and $q_3$ of the form $ZIIZ$. To study parameter uncertainty we simply mitigated one model by its perturbed version. However, because the model violation in this case is out-of-model, it is important that we also simulate the learning procedure, and we do so in the infinite shot limit to remove model-parameter uncertainty. This is because we anticipate that the learning procedure will adjust the in-model parameters to account for the out-of-model error. For our example, if the only error term was the out-of-model $ZIIZ$ with rate $\lambda_{ZIIZ}$, then we anticipate the learning procedure to produce a model with local error terms $IIIZ$ and $ZIII$ with rates $\lambda_{IIIZ}=\lambda_{ZIII}=\lambda_{ZIIZ}$. Empirically, this is also what we find when we start with an error channel with additional in-model terms.

\begin{figure*}[t!]
    \centering
    \includegraphics[width=2\columnwidth]{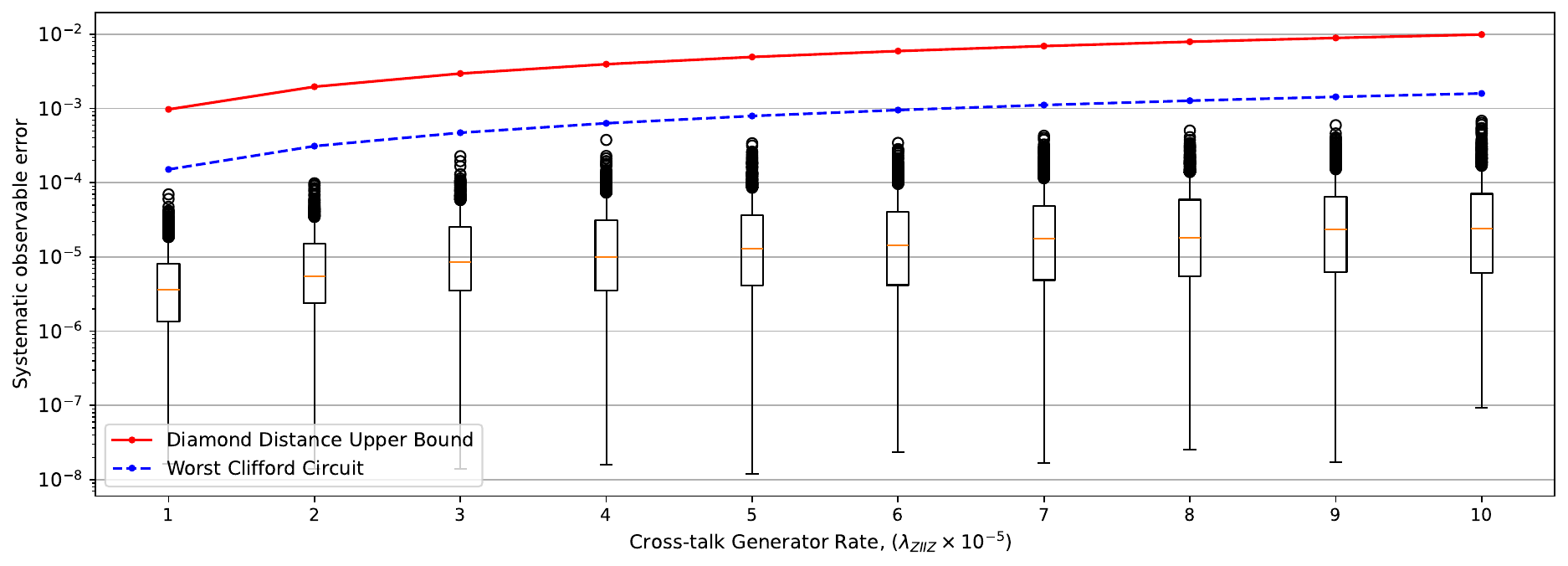}
    \caption{{\bf Out-of-model error simulations.} 
    Box and whisker plots of the distribution of systematic error after mitigation, $\delta_o$, for increasing out-of-model error $\lambda_{ZIIZ}$. The fixed in-model component of the actual error channel has parameters $\lambda_k$ drawn uniformly at random from $\left[10^{-5},10^{-4}\right]$, which is chosen to match current expectations for experimental gate fidelities. 
    The circuit structure is that shown in Fig.~\ref{fig:circs} (lower panel) repeated to a depth of 4. For the fixed initial state $\ket{0}^{\otimes 4}$, we sample 1000 random circuits and for each measure a random $4$-qubit Pauli observable. The red solid line plots the diamond distance upper bound to the systematic error of any circuit, the smaller of $\delta_\gamma$ of Eq.~\eqref{eqn:pecbound_dne} or $\delta_2$ of Eq.~\eqref{eqn:pecbound_2norm}. The blue dashed line plots the worst-case Clifford circuit systematic error of Eq.~\eqref{eqn:pecbound_wce}.}
    \label{fig:xtalk}
\end{figure*}

Figure \ref{fig:xtalk} shows the results of our learning and mitigation simulations as a function of the strength of the cross-talk error. Starting with an error channel consisting of in-model terms $\{\lambda_k\}$ drawn uniformly at random plus the cross-talk term with rate $\lambda_{ZIIZ}$ we simulate Pauli learning to fit a pairwise-local model. We then use that model to simulate mitigation of the actual error channel for a circuit consisting of 4 repetitions of the structure shown in the lower panel of Fig.~\ref{fig:circs}. As expected, the upper bound and worst-case Clifford systematic error grow with increasing cross-talk. This type of model violation shows a particularly large number of outliers clustering near the worst-Clifford circuit, which we believe is indicative of the fact that only half of the error channel's Pauli fidelities are impacted by the cross-talk term, such that the distribution for systematic error is not very normal.

We have included the random (but zero-uncertainty) in-model error terms in-order to highlight the drastic impact they have on the separation between the diamond distance upper bound and the worst-case Clifford circuit, even when simulating model learning. Without the random in-model error, as previously mentioned the Pauli learning trivially returns a model with nonzero $IIIZ$ and $ZIII$ rates $\lambda_{IIIZ}=\lambda_{ZIII}=\lambda_{ZIIZ}$. In this case, the mitigated map $\mathcal{V}$ has a generalized process fidelity $\nu_0 > 1$, which can be understood by the fact that all the Pauli fidelity ratios are greater or equal to one, as any Pauli that does not commute with $ZIIZ$ also does not commute with one of $IIIZ$ or $ZIII$, but there are Pauli operators which commute with $ZIIZ$ but not $IIIZ$ or $ZIII$.  As a result, Eq.~\eqref{eqn:pecbound_dne} becomes $\delta_\gamma = \gamma^l - 1$ for $l$ repetitions of the circuit in the lower panel of Fig.~\ref{fig:circs}, with $\gamma = e^{4\lambda_{ZIIZ}}$. The worst-case Clifford circuit also achieves a systematic error of $e^{4l\lambda_{ZIIZ}} - 1$, such that the $\delta_\gamma$ upper bound is tight and saturated by a Clifford circuit. We believe that with the random in-model error terms the actual worst-case systematic error is close to that achieved by the worst-case Clifford circuit, but our diamond norm estimation is no longer accurate enough to capture this.

\begin{figure*}[t!]
    \centering
    \includegraphics[width=2\columnwidth]{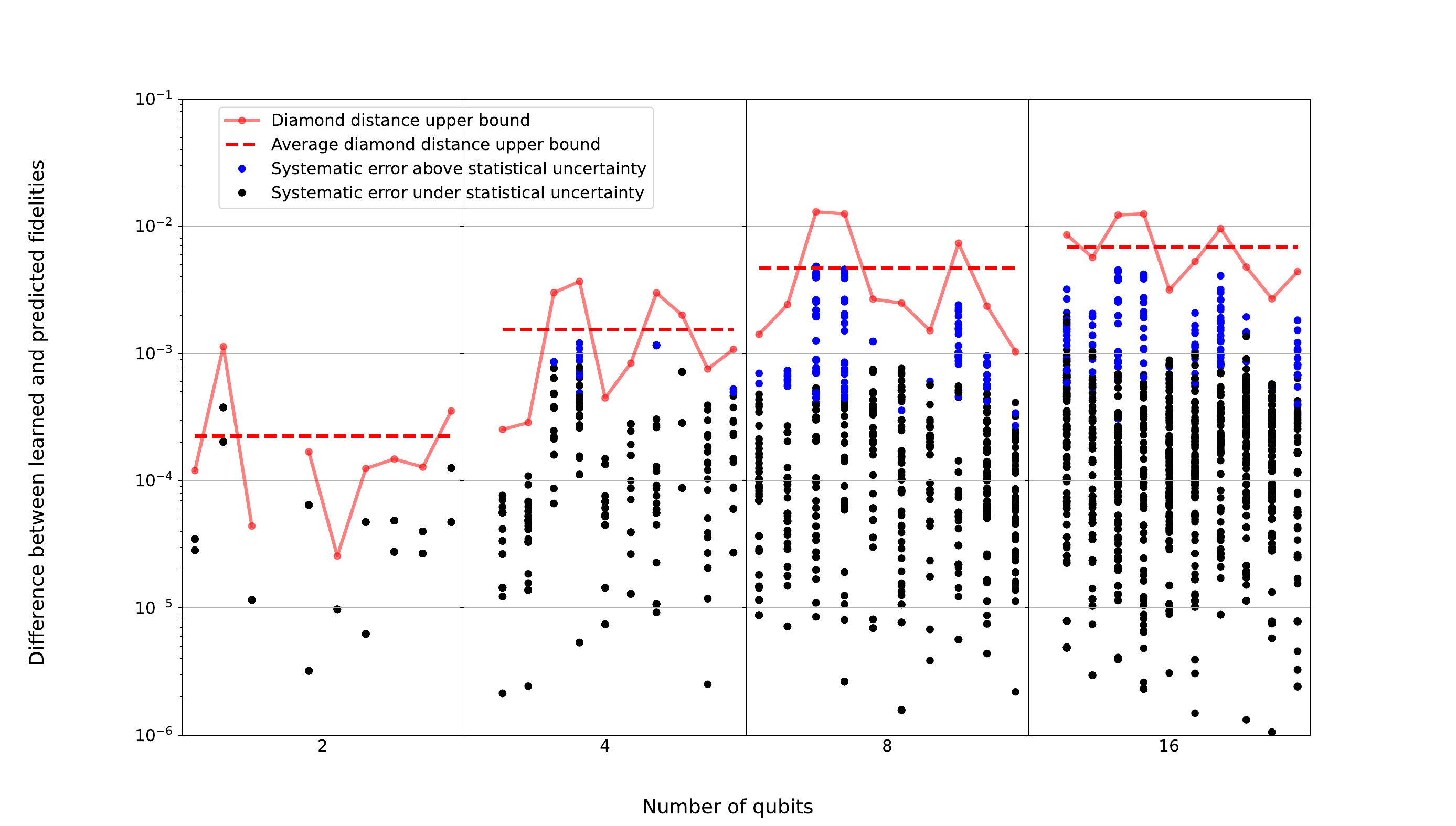}
    \caption{{\bf Experiments on {ibm Kyiv}.} 
    Model violation analysis of Pauli-learning data from connected qubit lines consisting of $2$, $4$, $8$, or $16$ qubits, with 10 sets for each qubit number. Scatter plots show the distribution of systematic error after simulated mitigation, $\abs{1 - f^{\rm meas}_P/f^{\rm mod}_P}$, for all Pauli operators $P$ measured in Pauli-learning. Points highlighted in blue have systematic error larger than the estimated statistical learning uncertainty, see Appendix \ref{sec:app_learnunc}. The red solid line plots for each qubit set the diamond distance upper bound to the systematic error of any circuit, the smaller of $\delta_\gamma$ of Eq.~\eqref{eqn:pecbound_dne} or $\delta_2$ of Eq.~\eqref{eqn:pecbound_2norm}, and the dashed red line indicates the average upper bound over qubit sets.
    The circuit consists of even connection {\tt CNOT} gates along the line of qubits. For Pauli-learning we use circuit depths of $[0,2,4,16,32,64]$, with $100$ Pauli-randomized twirl instances per learning circuit at each depth, and $200$ shots of each randomization. For one 2-qubit set the upper bound was very close to zero, and this data set has been excluded from the plot, leading to the discontinuous upper bound curve.}
    \label{fig:kyiv_data}
\end{figure*}

\section{Experimental Implementation on an IBM Quantum Processor}
\label{sec:exp}

\begin{figure}[t!]
    \centering
    \includegraphics[width=\columnwidth]{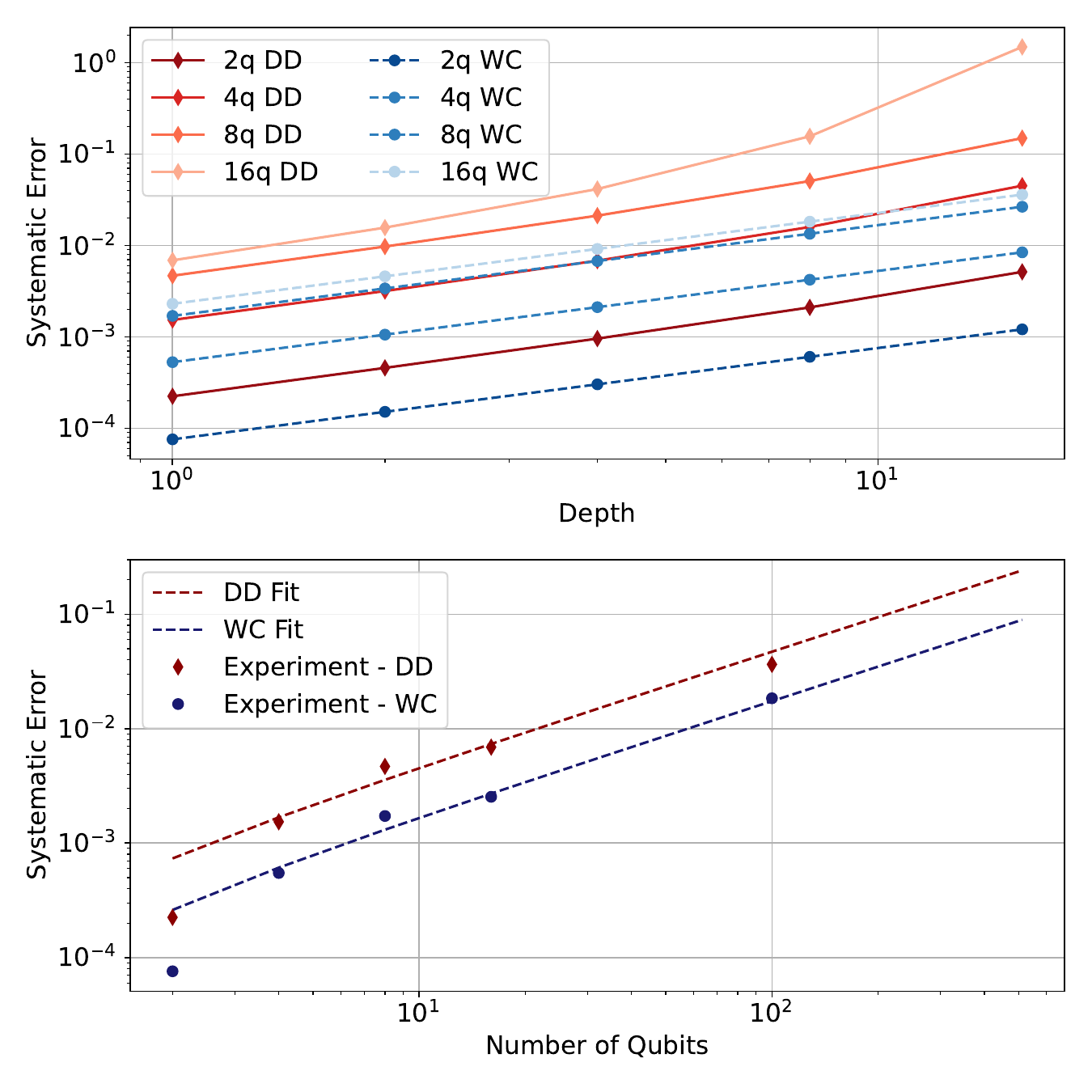}
    \caption{Scaling of the diamond distance upper bound (DD), the smaller of $\delta_\gamma$ of Eq.~\eqref{eqn:pecbound_dne} or $\delta_2$ of Eq.~\eqref{eqn:pecbound_2norm}, and the worst-case Clifford circuit of Eq.~\eqref{eqn:pecbound_wce} (WC) as a function of {\bf (Upper panel)} circuit depth, or {\bf (Lower panel)} qubit number at fixed depth of one. The data points at each qubit number are averaged over the 10 sets of data taken on distinct qubit sets of {\tt ibm Kyiv}.}
    \label{fig:kyiv_extrapolation}
\end{figure}

Now we turn to testing our methodology on the quantum processor {\tt ibm Kyiv}, a cross-resonance device with 127 superconducting qubits, available through the IBM Quantum Platform. In the previous section we used mitigated performance of non-Clifford circuits to compare against our upper bounds, but on a real processor additional experimental complexities (e.g.~finite sampling of mitigation experiments, or error channel drift between the Pauli learning and mitigation) make it difficult to isolate model violation in such comparisons. Instead, to get a sense of the model violation present in current generation processors, we run Pauli-learning experiments on a single layer of {\tt CNOT} gates across the even ($0-1$, $2-3$, etc.) links of a line of qubits. We then compute the diamond norm upper bounds for a depth-one circuit consisting of one instance of this layer, using $f_P^{\rm meas}$ extracted from the curve fits in Pauli-learning, and $f_P^{\rm mod}$ predicted by the model of Eq.~\eqref{eqn:PLmodel}, with associated generator rates $\vec{\lambda}$ learnt from the set of $f_P^{\rm meas}$. We compare this to the performance of the infinite-shot limit of mitigation of depth-one Clifford circuits, which we call ``simulated'' mitigation. This simulated mitigation can be computed by taking the ratio of the measured and model-predicted Pauli fidelities.

The results are shown in Fig.~\ref{fig:kyiv_data}, for 10 sets of each qubit number ($2$, $4$, $8$, or $16$) on {\tt ibm Kyiv}. For each qubit set, the diamond norm upper bound (smaller of $\delta_\gamma$ or $\delta_2$) is indicated with a red point and solid line. As expected, the (black/blue) scatter points showing the systematic error in mitigating a measured Pauli fidelity with its corresponding model prediction are all below the diamond distance upper bound. For these scatter points we can determine if the systematic error is larger than expected from statistical sources by computing the uncertainty in our model prediction $f^{\rm mod}_P$. Our methodology for this is described in Appendix \ref{sec:app_learnunc}. The scatter points coloured blue indicated systematic error larger than one-$\sigma$ of statistical uncertainty. As can be seen in Fig.~\ref{fig:kyiv_data}, for 4 qubits and beyond the majority of the high flier points are larger than one-$\sigma$ of statistical uncertainty in the learning, pointing to potential sources such as out-of-model error or non-Markovianity. In Appendix \ref{app:degen} we investigate the potential that the high fliers can be explained by the learning-degeneracy, and find that this is likely not the entire explanation.

The average upper bound over all qubit sets of a given qubit number is plotted as a dashed red line in Fig.~\ref{fig:kyiv_data}. We can similarly calculate the average upper bound for deeper circuits to understand how it scales with circuit depth. This is shown in the upper panel of Fig.~\ref{fig:kyiv_extrapolation}, along with the worst Clifford circuit for these larger circuit depths. The diamond distance upper bounds are guaranteed to grow at least linearly due to the sum over layers structure of Eq.~\eqref{eqn:PECgen}. The observed super-linear growth is due to the non-CP weighting factor discussed in section \ref{sec:PEC}. For deeper circuits the diamond distance upper bound can be quite pessimistic, and for that reason we also show the worst case Clifford circuit defined in Eq.~\eqref{eqn:pecbound_wce}. Given the form of Eq.~\eqref{eqn:pecbound_wce}, this should grow linearly in the low-error regime before turning over and approaching one asymptotically. Informed by the numerical simulations of section \ref{sec:numsims}, the order-of-magnitude lower worst-Clifford systematic error is a good heuristic upper bound for generic circuits.

Fig.~\ref{fig:kyiv_data} clearly elucidates that the systematic error due to model violation also grows as qubit-number increases. Fitting the data points to a line, we can extrapolate out to larger qubit set sizes, as shown in Fig.~\ref{fig:kyiv_extrapolation}. Similarly, we can fit the worst-Clifford circuit to a linear extrapolation, which is also shown. At a scale of 200 qubits, both predict a systematic error for a depth-one circuit of order $10^{-1}$. In order to build confidence in our prediction, we further ran an experiment with 100 qubits and find that the experimental systematic error matches our prediction (from the linear fit of 2, 4, 8, and 16 qubits) very well.  A linear extrapolation is valid in the low model violation regime when the source is either statistical or due to cross-talk, but is used here mostly as a ballpark estimate for the performance of systems scaling in width. The non-linear jump between 2 and 4 qubits may indicate the presence of out-of-model cross-talk. The 2-qubit model contains all possible interaction terms, but for $\geq 3$ qubits the model is incomplete as the only allowed interactions are those that respect the device connectivity.

\section{Comparing Models via PEC}
\label{sec:stability}

\begin{figure*}[t!]
    \centering
    \includegraphics[width=2\columnwidth]{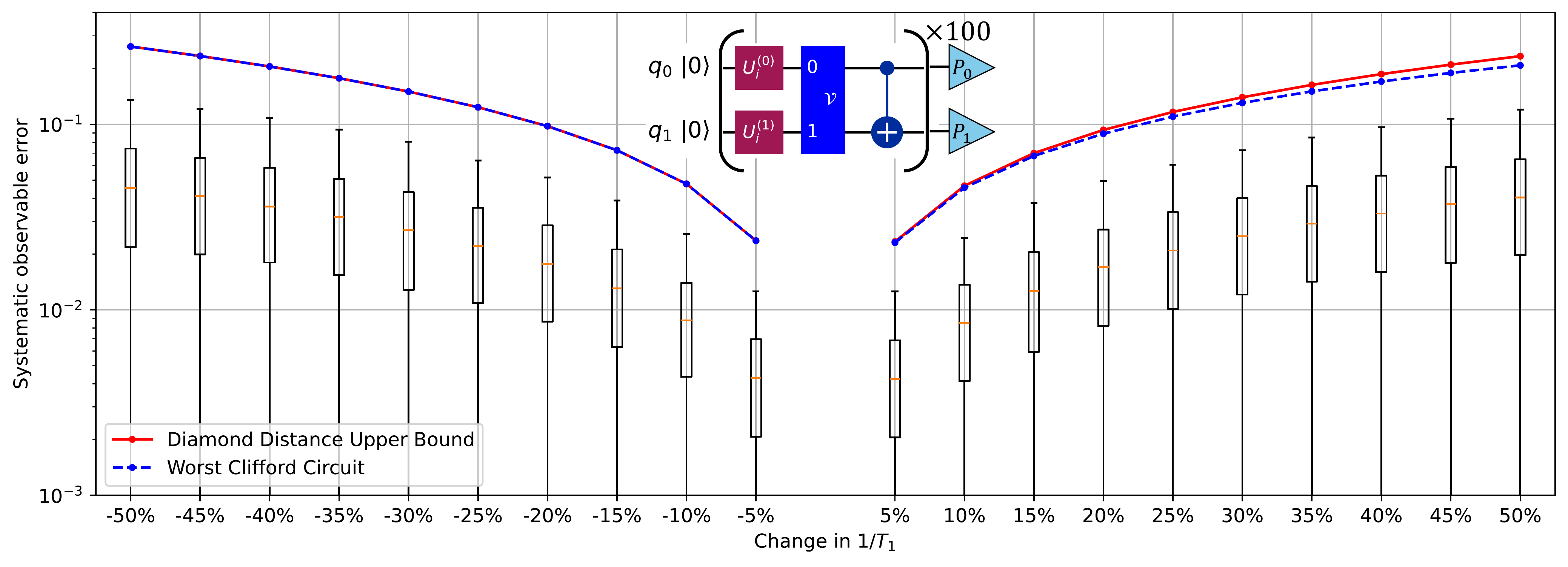}
    \caption{{\bf Coherence drift simulations.} Box and whisker plots of the distribution of systematic error after mitigation, $\delta_o$, as a function in the change in relaxation rate, $1/T_1$, for random circuits, random Pauli observables, and a fixed initial state $\ket{00}$. The coherence times of the model used for mitigation are $T_1 = 200\mu$s, $150\mu$s for qubit $0$ and $1$ respectively, with an assumed evolution time of $100$ ns. For each change in $1/T_1$ point, we sample 1000 circuits of the structure shown in the inset, and calculate the expectation value of all 15 non-trivial two-qubit Pauli operators ($\mathbb{I}\mathbb{I}$ is excluded). The red solid line plots the diamond distance upper bound to the systematic error of any circuit obtained via model comparison, $\delta_C$ of Eq.~\eqref{eqn:dnorm_comp}. The blue dashed line plots the worst-case Clifford circuit systematic error of Eq.~\eqref{eqn:pecbound_wce}. Inset shows the circuit structure for the simulations, where the $U_i^{(q)}$ gates are independently random at each layer. Lower quartiles of the box and whisker plots are cut from view to focus on the more important information at higher systematic error.}
    \label{fig:T1drift}
\end{figure*}

The model violation methodology presented so far has focused on understanding how well a learnt model is able to mitigate an underlying fixed error channel, by quantifying how well the model describes the true error. Reverting cause and effect, model violation can also be used to directly compare models by estimating the worst case error if one error model was used to mitigate the other. Such comparisons are useful to, for example, test the stability of error models over time, when error parameters are expected to drift or fluctuate, or to test the impact in uncertainty induced by the model fitting procedure.

For this analysis, because we have complete information about the error channels for both models, we can directly calculate upper bounds to the diamond distances and norms in Eq.~\eqref{eqn:PECgen} from the model parameters. Following the methodology of \cite{Berg2023}, for each layer the mitigation map $\mathcal{V}_k$ can be decomposed into a composition of Pauli Stochastish channels each with a single Pauli operator, $\mathcal{V}_k = \prod_m \mathcal{W}^{(k)}_m$. These take the form $\mathcal{W}^{(k)}_m = e^{2\hat{\lambda}_m}\mathcal{P}_m\hat{\mathcal{P}}^{-1}_m$, which is the composition of a single-term Pauli-stochastic channel $\mathcal{P}_m$ for one model, given by
\begin{align}
    \mathcal{P}_m(\rho) = w_m\rho + (1 - w_m)P_m\rho P_m,
\end{align}
with the inverse of the corresponding channel for the other model
\begin{align}
    \hat{\mathcal{P}}^{-1}_m(\rho) = \hat{w}_m\rho + (\hat{w}_m -1)P_m\rho P_m.
\end{align}
Here $w_m = (1+e^{-2\lambda_m})/2$, and $\hat{w}_k$ is defined analogously.

Applying the same approach we did to multi-layer circuits to the multi-component description of a single layer's mitigated map we can write the diamond distance for each layer as
\begin{align}
    \dnorm{\mathcal{I} - \mathcal{V}_k} \leq \sum_m \dnorm{\mathcal{I} - \mathcal{W}^{(k)}_m}\prod_n^{m-1}\dnorm{\mathcal{W}^{(k)}_n},
\end{align}
and the overall upper bound for observable error comparing the models is then given by
\begin{align}
    \nonumber \delta_o &\leq \sum_{k} \left(\sum_m \dnorm{\mathcal{I} - \mathcal{W}^{(k)}_m}\prod_n^{m-1}\dnorm{\mathcal{W}^{(k)}_n}\right) \prod_j^{k-1}\dnorm{\mathcal{V}_j} \\
    &\equiv \delta_C. \label{eqn:dnorm_comp}
\end{align}
Every term in this expression can be evaluated exactly. The expressions for $\dnorm{\mathcal{V}_j}$ and $\dnorm{\mathcal{W}^{(k)}_n}$ are derived in Eq.~\eqref{eqn:mod_dnorm} of Appendix \ref{app:invdnorm}, and $\dnorm{\mathcal{I} - \mathcal{W}^{(k)}_m}$ follows from that derivation. They are given by
\begin{align}
    &\nonumber\dnorm{\mathcal{I} - \mathcal{W}^{(k)}_m} = \abs{1 - \frac{e^{-2\lambda^{(k)}_m}}{e^{-2\hat{\lambda}^{(k)}_m}}}, \\
    &\nonumber \dnorm{\mathcal{W}^{(k)}_n} = \frac{1}{2}\left( \abs{1 + \frac{e^{-2\lambda^{(k)}_m}}{e^{-2\hat{\lambda}^{(k)}_m}}} +  \abs{1 - \frac{e^{-2\lambda^{(k)}_m}}{e^{-2\hat{\lambda}^{(k)}_m}}}\right), \\
    &\nonumber\dnorm{\mathcal{V}_k} \leq \prod_{k} \dnorm{\mathcal{W}^{(k)}_n},
\end{align}
for a Pauli-stochastic model with error rates $\lambda^{(k)}$ at layer $k$ being mitigated by a model with parameters $\hat{\lambda}^{(k)}$.

The model comparison described here is insensitive to the model violation due to inconsistency between each model and the data from which it is fit, which has been the focus of the rest of this work. As such, this comparison isolates the model violation due solely to changing model parameters. However, changing model parameters do not necessarily mean a change in the underlying physical error channel. Any realistic learning experiment takes only finite samples, which induces uncertainty in the learnt model parameters. This implies that the learning procedure applied to the same error channel twice may not produce the same model in both instances of learning. Moreover, differing models for the same error channel can result from a numerical method for calculating the model parameters that is nondeterministic (e.g.~due to optimization with different initial seeds). 

Thus, in practice one must be careful when interpreting the results of model comparison used for error channel fluctuation analysis, and independently estimate the potential contribution due to model parameter uncertainty, as we have done for our experimental data in section \ref{sec:exp}. The model uncertainty simulations of section \ref{sec:uncertainty} were analyzed in a way to mimic the experimental situation, where we do not have access to the full description of the actual error channel. However, since in simulation we do have both models, they can also be analyzed as a comparison between models that differ due to statistical fluctuations. In Fig.~\ref{fig:uncertainty_sup} of Appendix \ref{sec:cohdrift}  we show a reproduction of Fig.~\ref{fig:uncertianty} including a new upper bound calculated via the model comparison methodology. This also functions as a comparison between the three methodologies for approximating the diamond distance used in this work, via their respective upper bounds $\delta_\gamma$, $\delta_2$, and $\delta_C$.

\subsection{Coherence time drift}

Returning to our original intended use case for model comparison, we simulate a two-qubit system where the relaxation time $T_1$ of each qubit drifts on a slow timescale, such that the $T_1$ during model learning differs from that of the mitigation experiments. We use a simple test circuit (see inset of Fig.~\ref{fig:T1drift}), consisting of $l=100$ repetitions of a layer structure of independently random single-qubit gates followed by a {\tt CNOT} gate (the single-qubit gates are random per layer and per qubit), with a single instance of the mitigated channel, $\mathcal{V} = \mathcal{E}\hat{\mathcal{E}}^{-1}$, before every {\tt CNOT}. 

In Fig.~\ref{fig:T1drift}, we show the results of these simulations as a function of how much the relaxation rate $1/T_1$ of each qubit changes between the $\hat{\mathcal{E}}$ model we assume is the learnt model used for mitigation, and the $\mathcal{E}$ model we use to represent the actual error channel during mitigation experiments. As expected, the systematic error increases for larger changes in $1/T_1$, and the distribution of systematic error is well upper-bounded by the diamond distance upper bound calculated with the model comparison methodology, $\delta_C$. Extrapolating linearly in qubit number we estimate an up to $<5\%$ change in $1/T_1$ can be tolerated for a 50 qubit, depth 100 circuit to have systematic error below $10^{-1}$ in the worst case where all changes in $T_1$ align. More realistically, the fluctuations will not all align and a larger deviation in $T_1$ will be tolerable.

Of note is the fact that the worst-case Clifford circuit saturates the upper bound, $\delta_C$, when $1/T_1$ decreases between learning and mitigation (where the mitigation over-corrects the error), and very nearly saturates it when $1/T_1$ increases. As this error model is simple, Eq.~\eqref{eqn:dnorm_comp} can be calculated analytically, and one can show that for any number of qubits, if the relaxation rates of all the qubits decrease, then the worst-case Clifford circuit exactly equals the upper bound $\delta_C$ of Eq.~\eqref{eqn:dnorm_comp}. In this case, the worst case Clifford circuit corresponds to back-propagating the all Pauli-$Z$ observable. See Appendix \ref{sec:cohdrift} for further details.

\section{Conclusion}
\label{sec:conc}

In this work, we have presented several approaches to bounding the systematic error in error mitigation due to inaccuracies in the learnt noise model. Our methodology efficiently estimates upper bounds to systematic error purely from the existing experimental data collected during model learning at no additional experimental cost, and we have demonstrated the application of our methods on an IBM Quantum device. These upper bounds are based on constructions involving the diamond distance between the ideal and mitigated circuits, which is known to be a pessimistic measure of performance, in that the worst-case it measures may be far from average. However, our numerical simulations give evidence that the worst-case Clifford circuit is typically also a good heuristic measure of worst-case systematic error, though we have also proven that for some circuit structures worse non-Clifford circuits exist. Finally, we have shown that our methodology can be extended to model comparison, such as for analysis of noise fluctuation and drift. An important consideration is that our methodology takes the measured Pauli fidelities as ground truth, and as such is insensitive to systematic bias that affects all measured Pauli fidelities in the same way. Thus, our protocols must be supplemented for checks of such systematic bias, though we anticipate this to be a rare and pathological situation in a well calibrated device.

There are two qualitative sources of looseness in our upper bounds. Firstly, we upper bound the diamond norm with expressions that can be computed efficiently from experimental data. The magnitude by which this overestimates the diamond norm depends on the specifics of the mitigated channel. In some cases, such as the generalized depolarizing channel or mitigated amplitude damping, our upper bounds are exactly the diamond norm. We hope that with further research these upper bounds can be brought closer to the diamond norm for more general cases. Secondly, our target is the systematic error in a Pauli observable expectation value, and such observables may not be able to achieve as large a systematic error as predicted by the diamond norm. For example, the diamond distance to identity of the generalized depolarizing channel is a factor of $2 (4^n-1)/4^n$ greater than what can be achieved for any Pauli observable, but for an amplitude damping channel the diamond distance and worst-case Clifford circuit can coincide. Exploring the nature of this fundamental gap for more general mitigated channels is another topic for future work.

Our choice of the diamond norm follows from our choice to upper bound systematic observable error in-terms of the trace-distance of the circuit output states and the spectral norm of the observable. The trace-distance is then naturally upper bounded by the diamond distance in a way that is initial-state independent. An alternate option is to instead upper bound in terms of the spectral-norm induced distance between circuit output states and the trace-norm of the observable, as was done in Ref.~\cite{Henao23}. The advantage of this approach is that the spectral-norm induced distance is simple to calculate for Pauli Stochastish channels, and amounts to the largest difference between Pauli fidelities. However, the trace-norm of a Pauli observable grows exponentially with system size. As such, while this approach may be useful for a small number of qubits, it quickly becomes a looser upper bound than that given by our approach using the diamond distance.

The discussion in this work has focused on how to use model violation to quantify the systematic error in a mitigated observable due to an incorrect model; however, the methodology presented can alternatively be understand as a protocol to test the quality of an error model. The error model quality is measured via an operationally meaningful metric, the upper bound to $\delta_o$, that quantifies how poorly the model could fail at PEC error mitigation. Such an operationally defined metric is useful as it gives a clear prescription for an experimental test for verifying its measure of quality, which in this case is running PEC error mitigation with the learnt model. We leave as a question for future work how this metric correlates with the potential success or failure of other applications of learnt error models, including other error mitigation schemes, and noise-aware decoding in quantum error correction \cite{Heim16}.

\acknowledgments

We thank David Layden, Ewout van den Berg, Moein Malekakhlagh, and Abhinav Kandala for careful reading of the manuscript and enlightening feedback. The code used to collect the experimental data was developed by Ewout van den Berg, and we thank him for his support. Research was partially sponsored by the Army Research Office and was accomplished under Grant Number W911NF-21-1-0002. The views and conclusions contained in this document are those of the authors and should not be interpreted as representing the official policies, either expressed or implied, of the Army Research Office or the U.S. Government. The U.S. Government is authorized to reproduce and distribute reprints for Government purposes notwithstanding any copyright notation herein.

\appendix

\section{Further Details of Diamond Norm Upper Bounds}
\label{app:bounds}

\subsection{Diamond Norm Bounds for Multi-Layer Circuits}
\label{app:DNbounds}

Using the nomenclature from the main text with $\mathcal{V} = \mathcal{E}\hat{\mathcal{E}}^{-1}$, we have that $\delta_o$ for an arbitrary layered circuit is upper bounded by
\begin{align}
    \delta_o \leq \dnorm{\prod_{j=1}\mathcal{U}_j - \prod_{j=1}\mathcal{U}_j\mathcal{V}_j},
\end{align}
where $\mathcal{U}_j$ is the ideal unitary for each layer (merging easy layers into adjacent hard-layers) such that $\mathcal{C} = \prod_j\mathcal{U}_j$. Using sub-multiplicativity of the diamond norm and the fact that $\dnorm{\Lambda}=1$ for $\Lambda$ a CPTP channel, we obtain
\begin{align}
    \delta_o &\leq \dnorm{\prod_{j=2}\mathcal{U}_j - \mathcal{V}_1\prod_{j=2}\mathcal{U}_{j}\mathcal{V}_j} \\
    &\leq \dnorm{\mathcal{I} - \mathcal{V}_1} + \dnorm{\mathcal{V}_1}\dnorm{\prod_{j=2}\mathcal{U}_j - \prod_{j=2}\mathcal{U}_j\mathcal{V}_j}.
\end{align}
Repeating this procedure for the entire circuit we arrive at the expression in the main text
\begin{align}
    \delta_o \leq \sum^l_j \dnorm{\mathcal{I} - \mathcal{V}_j}\prod_k^{j-1}\dnorm{\mathcal{V}_k}. \label{eqn:dnormapp}
\end{align}

\subsection{Diamond Norm of a Pauli Stochastish Channel}
\label{app:dnormPS}

Following Ref.~\cite{Watrous13}, we write a Stinespring dilation of a Pauli Stochastish channel $\mathcal{V}$ as
\begin{align}
    &\mathcal{V}(\rho) = {\rm Tr}_{2}\left[A\rho B^\dagger\right] \\
    &A = B^\dagger = \sum_i \sqrt{\nu_i} P_i \otimes \ketbra{i},
\end{align}
where the auxiliary Hilbert space has dimension at most $4^n$. As shown in Ref.~\cite{Watrous13}, the diamond norm of $\mathcal{V}$ is equal to the maximum output fidelity of the maps $\Psi_A(\rho) ={\rm Tr}_{1}\left[A\rho A^\dagger\right] $ and  $\Psi_B(\rho) ={\rm Tr}_{1}\left[B\rho B^\dagger\right] $. Conveniently, we have that
\begin{align}
    \nonumber\Psi_A(\rho) &= {\rm Tr}_{1}\left[A\rho A^\dagger\right] = \sum_i \abs{\nu_i} {\rm Tr}\left[P_i\rho P_i\right]\otimes \ketbra{i} \\&= \sum_i \abs{\nu_i}\ketbra{i} = \Psi_B(\rho),
\end{align}
such that the output of both maps is independent of the input. From this we obtain that
\begin{align}
    \nonumber\dnorm{\mathcal{V}} &= \max_{\rho_A, \rho_B}F\left(\Psi_A(\rho_A),\Psi_B(\rho_B)\right) \\
    \nonumber&= F\left(\sum_i \abs{\nu_i}\ketbra{i},\sum_i \abs{\nu_i}\ketbra{i}\right) \\
    \nonumber&=\abs{\abs{\sqrt{\sum_i \abs{\nu_i}\ketbra{i}}\sqrt{\sum_i \abs{\nu_i}\ketbra{i}}}}_1 \\
    &= \abs{\abs{\sum_i \abs{\nu_i}\ketbra{i}}}_1 
    = \sum_i \abs{\nu_i}.
\end{align}

\subsection{Diamond Norm of the Mitigated Channel in PEC}
\label{app:invdnorm}

We can write the actual error channel $\mathcal{E}$ as a product of Pauli-stochastic channels
\begin{align}
    \mathcal{E} = \prod_k\mathcal{P}_k,
\end{align}
where
\begin{align}
    \mathcal{P}_k(\rho) = w_k\rho + (1 - w_k)P_k\rho P_k,
\end{align}
with $w_k = (1+e^{-2\lambda_k})/2$ \cite{Berg2023}. Each $\lambda_k$ is the model coefficient for Pauli $P_k$ in the error generator $\mathcal{L}(\rho) = \sum_k\lambda_k\left(P_k \rho P_k -\rho\right)$ that defines the actual error channel $\mathcal{E} = e^{\mathcal{L}}$. The inverse channel in PEC is implemented as a product of single element Pauli Stochastish channels \cite{Berg2023}, given by
\begin{align}
    \hat{\mathcal{E}}^{-1} = \gamma\prod_k\hat{\mathcal{P}}^{-1}_k,
\end{align}
where
\begin{align}
    \hat{\mathcal{P}}^{-1}_k(\rho) = \hat{w}_k\rho + (\hat{w}_k -1)P_k\rho P_k,
\end{align}
with $\hat{w}_k = (1+e^{-2\hat{\lambda}_k})/2$. As before, each $\hat{\lambda}_k$ is the model coefficient for Pauli $P_k$ in the error generator $\hat{\mathcal{L}}(\rho) = \sum_k\hat{\lambda}_k\left(P_k \rho P_k -\rho\right)$ that defines the total channel $\hat{\mathcal{E}} = e^{\hat{\mathcal{L}}}$ for the learnt error model. Using the result of Appendix \ref{app:dnormPS}, each $\hat{\mathcal{P}}^{-1}_k$ has a diamond norm given by
\begin{align}
    \dnorm{\hat{\mathcal{P}}^{-1}_k} = \abs{\hat{w}_k} + \abs{\hat{w}_k -1} = 1
\end{align}
where we have used the fact that $0 < \hat{w}_k \leq 1$. Thus, we see that the diamond norm of the inverse channel is upper bounded by
\begin{align}
    \dnorm{\hat{\mathcal{E}}^{-1}} = \gamma\dnorm{\prod_k\hat{\mathcal{P}}^{-1}_k} \leq \gamma\prod_k\dnorm{\hat{\mathcal{P}}^{-1}_k} = \gamma,
\end{align}
where $\gamma = \exp\left(2\sum_k \hat{\lambda}_k\right)$ is the overhead as defined in \cite{Berg2023}. This result has been previously shown in \cite{Regula2021}.

Since all single Pauli error channels such as $\mathcal{P}_k$ commute, we can use the descriptions of $\hat{\mathcal{E}}^{-1}$ and $\mathcal{E}$ in this form to upper bound the diamond norm of the mitigated map $\mathcal{V} = \mathcal{E}\hat{\mathcal{E}}^{-1}$. Starting from
\begin{align}
   \nonumber \mathcal{V} &= \prod_k \mathcal{W}_k =  \gamma\prod_k \mathcal{P}_k\hat{\mathcal{P}}^{-1}_k \\&= \gamma\prod_k\left[\left(\hat{w}_k + w_k  - 1\right)\rho + \left(\hat{w}_k - w_k\right)P_k\rho P_k\right],
\end{align}
and using the definition of $\hat{w}_k$, $w_k$, and $\gamma$ we obtain an expression for the diamond norm
\begin{align}
    \nonumber\dnorm{\mathcal{V}} &\leq \gamma\prod_k \frac{1}{2}\left(e^{-2\hat{\lambda}_k} + e^{-2\lambda_k} + \abs{e^{-2\hat{\lambda}_k} - e^{-2\lambda_k}}\right)
    \\ \nonumber  &= \prod_k \frac{1}{2}\left(1 + \frac{e^{-2\lambda_k}}{e^{-2\hat{\lambda}_k}} + \abs{1 - \frac{e^{-2\lambda_k}}{e^{-2\hat{\lambda}_k}}}\right)
    \\ &= \prod_{\hat{\lambda}_k > \lambda_k} \frac{e^{-2\lambda_k}}{e^{-2\hat{\lambda}_k}}, \label{eqn:mod_dnorm}
\end{align}
where in the second line we absorbed $\gamma$ into the product of terms. We cannot directly measure the $\lambda_k$ of the true error model and so we cannot calculate this expression directly. As such, in the main text we upper bound the diamond norm of the mitigated channel by that of the inverse error channel, i.e.~$\gamma$. 

\subsection{Upper Bound via Diamond Norm and Process Fidelity}
\label{subsec:dnpf}

Since the maps $\mathcal{V}_k$ are all trace-preserving (their identity Pauli fidelity ratio is $r_0=1$) we can use a result of Ref.~\cite{Magesan12} to calculate their diamond distance to identity 
\begin{align}
    \nonumber\dnorm{\mathcal{I} - \mathcal{V}} &= \abs{1-\nu_0} + \sum_{j=1} \abs{\nu_j} \\
    &= \abs{1-\nu_0} + \dnorm{\mathcal{V}}-\abs{\nu_0}. \label{eqn:ddist1}
\end{align}
While $\nu_0 > 0$ (for $\nu_0$ as defined previously), we may have that $\nu_0 > 1$ since the mitigated map need not be CPTP. This leads to the simplification
\begin{align}
    \dnorm{\mathcal{I} - \mathcal{V}} = \left\{\begin{array}{cc}
       1 + \dnorm{\mathcal{V}} - 2\nu_0  &  \nu_0 \leq 1\\
       \dnorm{\mathcal{V}} - 1  &  \nu_0 > 1
    \end{array}\right. .\label{eqn:ddist}
\end{align}
To upper bound $\dnorm{\mathcal{V}}$ we use $\gamma$ as described earlier. We can estimate $\nu_0$ from the set of $r_k = f^{\rm meas}_{P_k}/f^{\rm mod}_{P_k}$ that we have collected during the learning procedure via Eq.~\eqref{eqn:FTnu}. Since we cannot measure all $r_k$, this necessarily means our upper bound is an approximate estimate. From the above we arrive at the upper bound for $\delta_o$ in Eq.~\eqref{eqn:pecbound_dne}.

\subsection{Upper Bound via Two-Norm}
\label{subsec:2norm}

Using the relationship between the vector one and two norms, we can upper bound the first line in Eq.~\eqref{eqn:ddist1} by 
\begin{align}
    \nonumber\dnorm{\mathcal{I} - \mathcal{V}} &= \abs{1-\nu_0} + \sum_{j=1} \abs{\nu_j} \\ &\leq \abs{1-\nu_0} + \sqrt{(4^n-1)\sum_{j=1}\abs{\nu_j}^2}, \label{eqn:ddist2norm}
\end{align}
and our goal is then to estimate the last term in Eq.~\eqref{eqn:ddist2norm}. Using Eq.~\eqref{eqn:FTnu}, we can write
\begin{align}
    &\nonumber\sqrt{(4^n-1)\sum_{j=1}\abs{\nu_j}^2} \\ &\nonumber= \frac{\sqrt{4^n-1}}{4^n}\sqrt{\sum_{j=1}\left(\sum_k \left(-1\right)^{\left<j,k\right>}r_k\right)^2} \\
    &= \frac{\sqrt{4^n-1}}{4^n}\sqrt{\sum_{j=1}\sum_{k,m} \left(-1\right)^{\left<j,k\right>}\left(-1\right)^{\left<j,m\right>}r_kr_m}.
\end{align}
For fixed $k$ each vector $\left(-1\right)^{\left<j,k\right>}$ is a row of the Walsh-Hadamard matrix of order $4^n$. This allows us to evaluate the expression
\begin{align}
    \sum_{j=1}\left(-1\right)^{\left<j,k\right>}\left(-1\right)^{\left<j,m\right>} = \left\{\begin{array}{cc}
       4^n - 1&  k = m\\
       - 1  &  k \neq m
    \end{array}\right.
\end{align}
as it represents the inner product of two rows of the Walsh-Hadamard matrix with their first (identity) element removed. Putting this all together, we obtain
\begin{align}
    \dnorm{\mathcal{I} - \mathcal{V}} &\leq \abs{1-\nu_0} \\&\nonumber+ \frac{4^n-1}{4^n}\sqrt{\sum_{k}r_k\left(r_k - \frac{1}{4^n-1}\sum_{m \neq k}r_m\right)}, 
\end{align}
which we insert into Eq.~\eqref{eqn:PECgen} to obtain $\delta_2$, the upper bound on the diamond distance of Eq.~\eqref{eqn:pecbound_2norm}.

\section{Clifford Circuits are Not the Worst Case for PEC Error Mitigation}
\label{app:Clif_WC}

As discussed in the main text, circuits and error models with $\delta_o$ larger than that for the worst-case Clifford circuit appear to be rare. To find a counter-example, we use a specific kind of non-Clifford circuit consisting of alternating layers of ``hard'' Clifford gates, and ``easy'' Pauli-rotation gates, where a Pauli-rotation gate for Pauli $P$ and angle $\theta$ is defined by the unitary $U(P,\theta) = \exp(-i\theta P/2)$. The mitigated circuit is thus
\begin{align}
    \mathcal{C}_{\mathbb{M}} = \prod_k \mathcal{U}(P_k,\theta_k)\mathcal{U}(C_k)\mathcal{V}_k,
\end{align}
where $\mathcal{U}(P_k,\theta_k)$ and $\mathcal{U}(C_k)$ are the process representations of the Pauli-rotation gate and Clifford gate at layer $k$ respectively. From this expression it is clear that if the vector of angles $\vec{\theta} = \vec{0}$ then the circuit is Clifford, and we choose $C_k$ such that it is the worst-case Clifford circuit.

As $\delta_o$ is now a function of $\vec{\theta}$, our task is then to understand the behaviour of $\delta_o$ in the vicinity of $\vec{\theta} = \vec{0}$. For simplicity, we will work with the signed version of $\delta_o$ defined by
\begin{align}
    \delta^{\pm}_o(\vec{\theta}) = {\rm Tr}\left[O\left(\mathcal{C}(\rho) - \mathcal{C}_{\mathbb{M}}(\rho)\right)\right], \label{eqn:signedd0}
\end{align}
and we will set all $P_k = P$ for some $P$ that does not commute with $O$. Further, we will assume a static error channel, $\mathcal{V}_k = \mathcal{V}~\forall k$, and pick $O$ to be the observable with the worst-case Pauli fidelity ratio, such that setting all $C_k$ to identity achieves the worst-case Clifford circuit. Our simplified mitigation circuit construction is thus
\begin{align}
    \mathcal{C}_{\mathbb{M}} = \prod_k \mathcal{U}(P,\theta_k)\mathcal{V}, \label{eqn:count_circ}
\end{align}
and we need only keep track of the Pauli operators $O$ and $PO$, since $\mathcal{U}^\dagger(P_k,\theta_k)O\mathcal{U}(P_k,\theta_k) = \cos(\theta_k)O + i\sin(\theta_k)PO$.

If $\delta^{\pm}_o(\vec{0})$ is an extremum point, then we can say that the worst-case Clifford circuit is a (local) extrenum of systematic error. To construct an example for which this is not the case, we use the Clifford perturbation theory (CPT) of Ref.~\cite{Begusic23} to express Eq.~\eqref{eqn:signedd0} as
\begin{align}
    \delta^{\pm}_o(\vec{\theta}) &= \sum^l_{k=0} i^k {\rm Tr}\left[\rho P^kO\right] E^{(k)}, \\
    \nonumber E^{(k)} &= \sum_{1\leq j_1 < ... < j_k}\left(1-r_{PO}^{n_{\vec{j}}}r_O^{l-n_{\vec{j}}}\right)  \\ 
    \nonumber &\times \prod_{m \in \{j_1,..,j_k\}} \sin(\theta_m)\prod_{m\notin \{j_1,..,j_k\}} \cos(\theta_m)
\end{align}
where $l$ is the total circuit depth, and $r_O$ ($r_{PO}$) is the Pauli fidelity ratio of the map $\mathcal{V}$ for the Pauli operator $O$ ($PO$). At each order $k$, only $E^{(k)}$ is a function of $\vec{\theta}$ and it enumerates by the appearance of $k$ terms of the form $\sin(\theta_m)$ all possible ways in a circuit of depth $l$ that there can be $k$ non-trivial actions of the Pauli-rotation gates, which result in the transformed operator $P^kO$. The factor $\left(1-r_{PO}^{n_{\vec{j}}}r_O^{l-n_{\vec{j}}}\right) $ accounts for the comparison between the ideal and mitigated circuits. For a given configuration of non-trivially acting Pauli-rotation gates at locations $m = \{j_1,..,j_k\}$, the number $n_{\vec{j}}$ counts how many times $PO$ is propagated through the circuit. For example, between $j_1$ and $j_2$ this occurs $j_2-j_1$ times, but between $j_2$ and $j_3$ this never happens, as only $O$ is propagated because the second action of $P$ at location $j_2$ returns the operator to $O$.

As also realized in Ref.~\cite{Mitarai22}, a nice feature of CPT is the ease in which it allows one to take derivatives of parameterized circuits with respect to the circuit parameters. In particular, the derivative of $E^{(k)}$ to order $n$ evaluated at $\vec{\theta} = \vec{0}$ will be zero for all $k>n$, owing to at least one extra $\sin(\theta_m)$ term. This greatly reduces the depth of $k$ that needs to be considered. Of relevance to us are the first and second derivatives evaluated at $\vec{\theta} = \vec{0}$.  We first note that since $\left\{P,O\right\} = 0$, and we have chosen $\rho$ to be an eigenstate of $O$ such that ${\rm Tr}\left[\rho O \right] = \pm 1$, then ${\rm Tr}\left[\rho PO \right] = 0$ and thus
\begin{align}
    \left.\frac{\partial \delta^{\pm}_o}{\partial \theta_m}\right|_{\vec{\theta} = \vec{0}} = 0~\forall m.
\end{align}
The second derivative is also straightforward to evaluate
\begin{align}
    &\left.\frac{\partial^2 \delta^{\pm}_o}{\partial \theta^2_m}\right|_{\vec{\theta} = \vec{0}} = -\left(1-r_O^l\right){\rm Tr}\left[\rho O \right], \label{eqn:hes1}  \\
    &\left.\frac{\partial^2 \delta^{\pm}_o}{\partial \theta_m\theta_a}\right|_{\vec{\theta} = \vec{0}} = -\left(1-r_{PO}^{\abs{m-a}}r_O^{l-\abs{m-a}}\right){\rm Tr}\left[\rho O \right]. \label{eqn:hes2} 
\end{align}
Together, these results imply that the worst-case Clifford circuit is indeed a critical point in $\vec{\theta}$-space, but whether or not it is a maximum, minimum or neither will depend on the Hessian matrix formed by the second derivatives.

We now explicitly construct (and verify) an example where the Hessian matrix has both positive and negative eigenvalues and the worst-case Clifford circuit is thus a saddle point in $\vec{\theta}$-space. This implies that there exist non-Clifford circuits with larger $\delta_o$ than the worst-case Clifford circuit, and we also explicitly construct such a circuit. Our example consists of a depth $25$ single qubit circuit with $\mathcal{V}$ defined by $r_Z=0.9$ and $r_{X} = r_Y = 1$, using a Pauli-rotation gate with $P=X$, an observable $O = Z$, and initial state $\rho$ that is the $+1$ eigenstate of $Z$. For this circuit, the $25$-dimensional Hessian matrix has one positive eigenvalue (the rest are strictly negative), which implies a single direction in $\vec{\theta}$-space for which the Clifford circuit is not a local maximum. 

\begin{figure}[t!]
    \centering
    \includegraphics[width=\columnwidth]{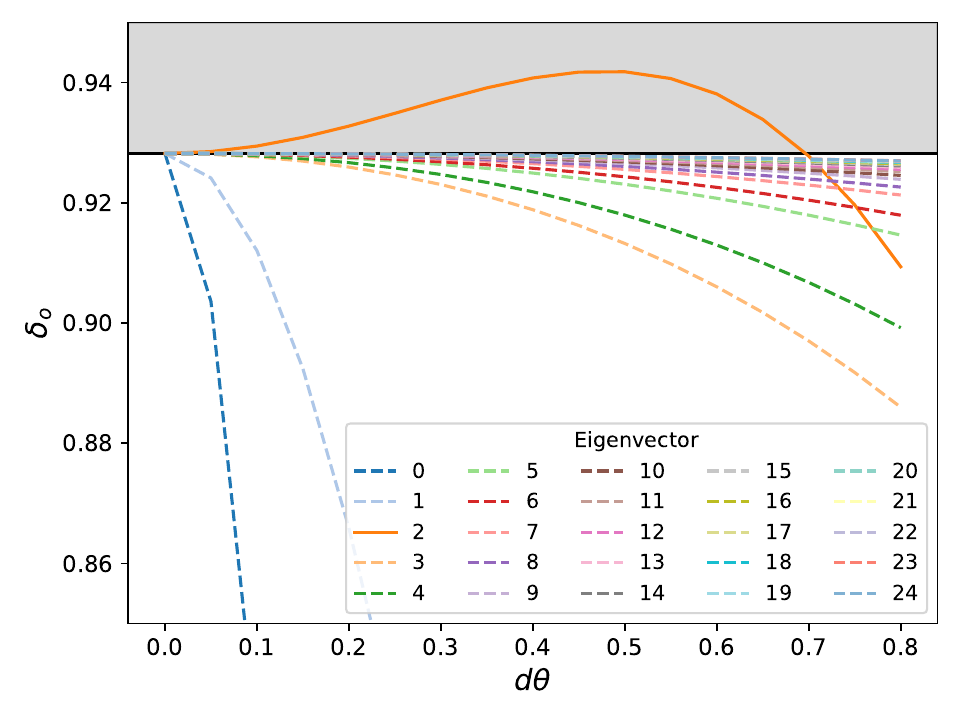}
    \caption{Systematic error $\delta_o = \abs{\delta^{\pm}_o}$ of the circuit construction of Eq.~\eqref{eqn:count_circ} as a function of distance $d\theta$ moved in $\vec{\theta}$-space along any of the 25 directions defined by the eigenvectors of the Hessian matrix of $\delta^{\pm}_o$ (see Eqs.~\eqref{eqn:hes1} and \eqref{eqn:hes2}). The black line indicates $\delta_o$ for the worst-case Clifford circuit, and the shaded region values of $\delta_o$ above this, which can only be achieved by non-Clifford circuits.}
    \label{fig:count_ex}
\end{figure}

To verify this we simulate this depth 25 circuit for $\vec{\theta} = d\theta \hat{\theta}_j$ where $\hat{\theta}_j$ is a normalized eigenvector of the Hessian matrix of $\delta^{\pm}_o$, and $d\theta$ is a dimensionless distance in $\vec{\theta}$-space. We repeat this simulation for all 25 eigenvectors, and plot $\delta_o = \abs{\delta^{\pm}_o}$ in Fig.~\ref{fig:count_ex}. The black line (at approximately $0.93$) is the systematic error due to model violation for the worst-case Clifford circuit, and values of $\delta_o$ in the shaded region above this line indicated systematic error for non-Clifford circuits that exceed that possible for a Clifford circuit. As can be seen, this occurs for only one eigenvector (eigenvector 2 shown with the orange solid line), and over a very small window of $\vec{\theta}$-space distance $d\theta$.

We note that for this circuit construction of a single-qubit and a fixed Pauli-rotation gate with independent angles at each layer, our numerical studies indicate there is a critical depth at which the Hessian matrix stops being negative definite, and a single positive eigenvalue emerges. This depth seems to depend on the value of the Pauli fidelity ratios, and was smaller than 25 for our chosen $r_Z = 0.9$. We used depth 25 in our example to highlight the effect and obtain a large difference in $\delta_o$ between the non-Clifford circuit and worst-case Clifford circuit. 

While the counter example we have presented here uses a mitigation map that is not a CPTP quantum channel, we have also found examples with the same circuit structure but a CPTP channel $\mathcal{V}$ where the worst-case circuit with largest $\delta_o$ is also non-Clifford. Thus, the onset of worse non-Clifford circuits at some critical depth does not require PEC error mitigation, but occurs even for physical error channels. Understanding the nature of the critical depth, i.e.~its dependence on error parameters and circuit structure, would be an interesting direction of future work characterizing the nature of noisy circuits with repeating layer structure. 

\section{Estimating model violation due to learning uncertainty}
\label{sec:app_learnunc}

Statistical uncertainty in the parameter values of a learnt model can be a significant source of model violation, but this is not a failure of the model ansatz. As such, it can be useful to independently estimate the learning uncertainty's effect on mitigated expectation values, such that one can potentially detect the presence of more nefarious sources of model violation such as out-of-model error. To that end, we develop here a procedure to estimate the systematic error in Clifford circuits due to learning uncertainty. 

For now, we assume that we are given a set of model coefficients $\{\lambda^{(k)}_j\}$ for each hard-layer $k$, and the covariance matrix $\boldsymbol{C}^{(k)}$ describing their correlation and uncertainty. We will discuss how to estimate this covariance matrix shortly. The noisy \textit{predicted} expectation value of a Clifford circuit with initial state $\rho$ and Pauli observable $O$ is simply given by $\left<O\right>_{\rm model} = f^{\rm model}_O\left<O\right>_{\rm ideal}$ where
\begin{align}
    f^{\rm model}_O = \prod_{k}^{l}f^{(k)}_{P_k},
\end{align}
is the product of the Pauli fidelity at each layer $k$ of the Pauli $P_k$ which results from back-propagating the observable $O$ through the last $l-k$ layers of the circuit. Each $f^{(k)}_{P_k}$ can be calculated from the model coefficients  $\{\lambda^{(k)}_j\}$ for each layer. Similarly, the noisy \textit{measured} expectation value of this Clifford circuit is given by $\left<O\right>_{\rm meas} = f^{\rm meas}_O\left<O\right>_{\rm ideal}$, where $f^{\rm meas}_O$ is defined in the same way as $f^{\rm model}_O$ but using the Pauli fidelities or model coefficients of the actual error channel (to which we do no have access).

PEC mitigation of a Clifford amounts to layer-by-layer application of the inverse Pauli fidelity $1/f^{(k)}_{P_k}$, which can be simplified to 
\begin{align}
    \left<O\right>_{\rm mitigated} = \frac{1}{f^{\rm model}_O}\left<O\right>_{\rm meas} = \frac{f^{\rm meas}_O}{f^{\rm model}_O}\left<O\right>_{\rm ideal}.
\end{align}
Calculating the systematic error for this circuit gives
\begin{align}
    \nonumber\delta_O = \abs{\left<O\right>_{\rm ideal} - \left<O\right>_{\rm mitigated}} = \abs{\left<O\right>_{\rm ideal}}\abs{1 - \frac{f^{\rm meas}_O}{f^{\rm model}_O}}.
\end{align}
Focusing on Clifford circuits where $\abs{\left<O\right>_{\rm ideal}} = 1$, in the absence of any other sources of model violation $\mathbb{E}\left[f^{\rm meas}_O/f^{\rm model}_O\right] = 1$. Thus, with one-$\sigma$ confidence $\delta_O$ will be less than the standard deviation of the ratio $f^{\rm meas}_O/f^{\rm model}_O$, which we define as $\sigma = \sigma\left[f^{\rm meas}_O/f^{\rm model}_O\right]$. Using standard propagation of error formulas assuming that $f^{\rm meas}_O$ and $f^{\rm model}_O$ are uncorrelated we have that
\begin{align}
    \sigma^2 \approx \left(\frac{f^{\rm meas}_O}{f^{\rm model}_O}\right)^2\left(\frac{\sigma^2\left[f^{\rm meas}_O\right]}{\left(f^{\rm meas}_O\right)^2} + \frac{\sigma^2\left[f^{\rm model}_O\right]}{\left(f^{\rm model}_O\right)^2}\right). \label{eqn:sigma}
\end{align}
The assumption of uncorrelated $f^{\rm meas}_O$ and $f^{\rm model}_O$ is likely valid if the model is fit from an independent data set from that used to calculated $\left<O\right>_{\rm meas}$. For the purposes of this manuscript, where we compare $f^{\rm model}_O$ to $f^{\rm meas}_O$ from the data set used to obtain $f^{\rm model}_O$, the two are likely somewhat positively correlated. This would reduce the variance of their ratio, $\sigma^2$, so we use the uncorrelated result of Eq.~\eqref{eqn:sigma} as a more stringent test of model violation beyond that due to statistical uncertainty.  

In order to estimate $\sigma$, we therefore need to estimate both $\sigma\left[f^{\rm meas}_O\right]$ and $\sigma\left[f^{\rm model}_O\right]$. There are several ways to obtain $\sigma\left[f^{\rm meas}_O\right]$, such as calculating it directly from the uncertainty in the Pauli-learning exponential fits, or estimating it via a bootstrap re-sampling of the data to generate a distribution of $f^{\rm meas}_O$ estimates. In our case we use the latter, as we will also need this bootstrap re-sampling to estimate the model prediction uncertainty $\sigma\left[f^{\rm model}_O\right]$.

\begin{figure*}[t!]
    \centering
    \includegraphics[width=2\columnwidth]{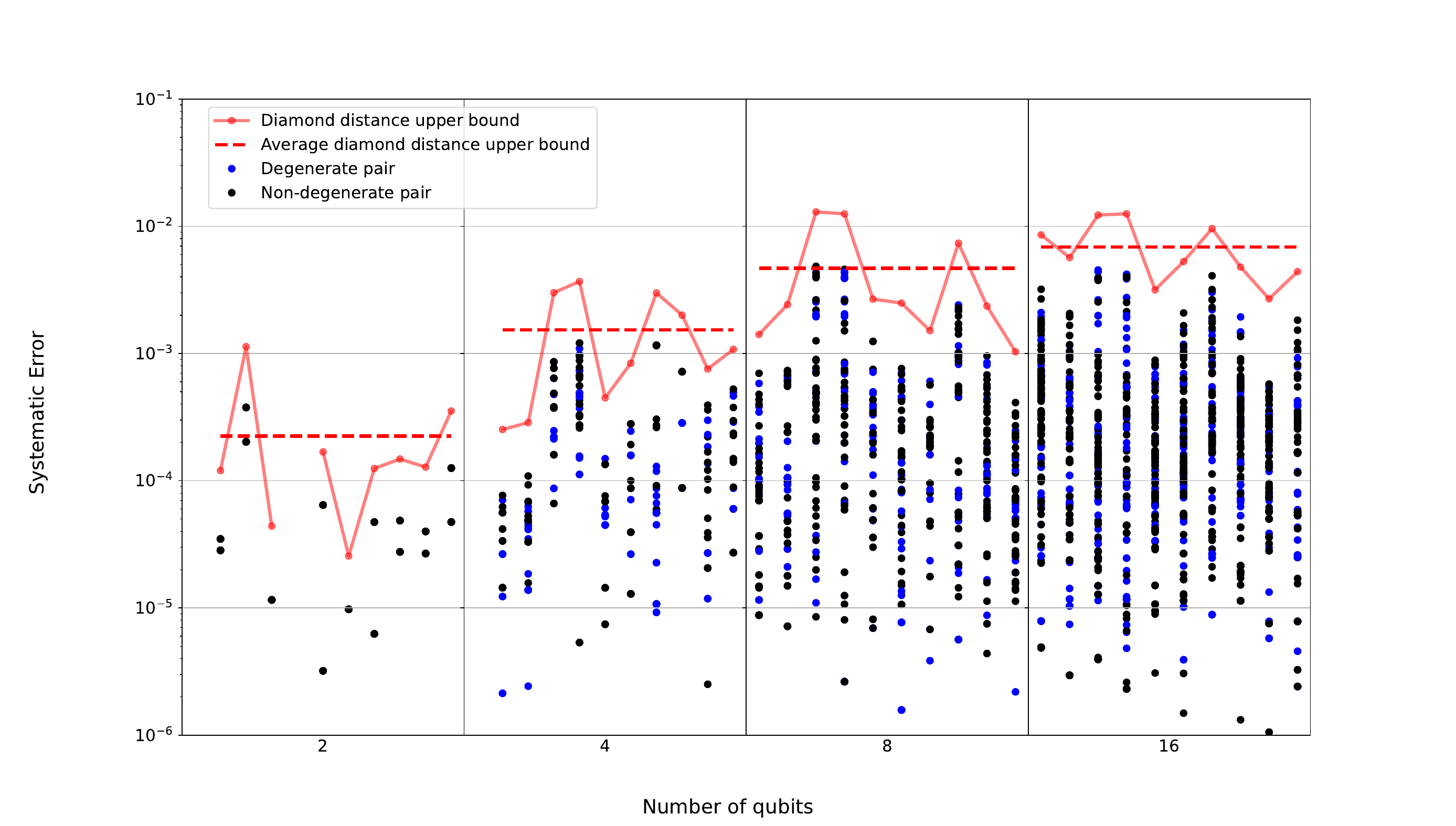}
    \caption{{\bf Experiments on {ibm Kyiv}.} 
    Model violation analysis of Pauli-learning data from connected qubit lines consisting of $2$, $4$, $8$, or $16$ qubits, with 10 sets for each qubit number. Scatter plots show the distribution of systematic error after simulated mitigation, $\abs{1 - f^{\rm meas}_P/f^{\rm mod}_P}$, for all Pauli operators $P$ measured in Pauli-learning. Blue scatter point correspond to Pauli fidelities that form part of a degenerate learning pair, with the degeneracy broken assuming $f_{P_1} = f_{P_2} = \sqrt{f_{P_1}f_{P_2}}$. Black scatter points correspond to Pauli fidelities that can be learnt uniquely with Pauli-learning. All else identical to Fig.~\ref{fig:kyiv_data}.}
    \label{fig:kyiv_data_degen}
\end{figure*}

The model prediction uncertainty $\sigma\left[f^{\rm model}_O\right]$ is due to uncertainty in the learnt model parameters $\{\lambda^{(k)}_j\}$, as described by their covariance matrix $\boldsymbol{C}^{(k)}$. To calculate $\sigma\left[f^{\rm model}_O\right]$, we must keep in mind that repeated layers will have the same model parameters, such that the full circuit prediction uncertainty is correlated across any repeated layers. It is thus easier to work with $f_O$ reformulated as the exponential of a sum over layers of a sum over all model coefficients at each layer
\begin{align}
   \nonumber f^{\rm model}_O &= \exp\left(-2\sum_k^l\sum_j \left[\vec{M}_{P_k}\right]_j \lambda^{(k)}_j \right) \\ &= \exp\left(-2\sum_{m=1}^{L}\lambda_m\right),
\end{align}
where $[\vec{M}_{P_k}]_j = 0$ if $P_k$ and $P_j$ commute and one if they anticommute, and in the second line we have combined the sum over layers and model coefficients into a sum over $L$ total model coefficients from different layers. For those $L$ coefficients we can create a circuit-level covariance matrix $\boldsymbol{C}$ from the relevant elements of each layer's covariance matrix $\boldsymbol{C}^{(k)}$, being careful to include not only the intra-layer covariances but also the inter-layer covariances between repeated layers. If no layers are repeated, then $\boldsymbol{C}$ has block diagonal structure, with each block corresponding to a different layer and formed by the restriction of $\boldsymbol{C}^{(k)}$ to the indices $j$ for which $[\vec{M}_{P_k}]_j = 1$.
Using approximate propagation of uncertainty formulas we can then express the uncertainty in $f^{\rm model}_O$ as
\begin{align}
    &\sigma\left[f^{\rm model}_O\right] \approx 2f_O^{\rm model}\sqrt{\sum_{m,n}^L \left[\boldsymbol{C}\right]_{m,n}}
\end{align}
This completes our estimate of Eq.~\eqref{eqn:sigma}, and it is $\sigma$ that we compare to the observed systematic error in Fig.~\ref{fig:kyiv_data} of the main text to determine which points show model violation beyond that expected due to statistical uncertainty.

What remains is to describe our procedure for estimating the covariance matrix of each layer. There are many ways to do this, and we choose to use a bootstrap re-sampling methodology. The Pauli-learning framework consists of a set of circuits chosen to estimate specific Pauli fidelities of a hard-layer \cite{Berg2023}, and each of these circuits defines a Pauli-randomized ensemble of circuits meant to Pauli twirl the error channel of the hard-layer. Each element of the Pauli-randomized ensembles is a binary outcome Clifford circuit, and from the experimental counts dictionary (or equivalently the expectation value) of this circuit we can define a corresponding binomial probability distribution for its shot-to-shot outcomes.

Our bootstrap re-sampling is two-fold. We first re-sample from the Pauli-randomized ensemble to obtain a new ensemble of Pauli-randomized circuits, and then for each of these circuits we sample from their corresponding binomial distribution to obtain a new counts dictionary (or equivalently a new expectation value). We can then process this re-sampled data set through the Pauli-learning and model fitting procedure: estimate the twirled expectation value of each circuit from the Pauli-randomized ensemble, fit the circuit decay curves to extract Pauli fidelities, and finally perform least squares optimization to obtain the model parameters \cite{Berg2023}. The output of this procedure is a new set of model parameters corresponding to the re-sampled data set. 

Repeating this re-sampling and model fitting many times, we obtain an ensemble of estimates for each model parameter, from which we can calculate a mean for each model parameter, $\{\lambda^{(k)}_j\}$, and the covariance matrix $\boldsymbol{C}^{(k)}$ describing their correlation and uncertainty. We then use these to estimate $\sigma\left[f^{\rm model}_O\right]$. Prior to model fitting, we also obtain an ensemble for each measured Pauli fidelity $f^{\rm model}_P$, which as mentioned previously we use to estimate $\sigma\left[f^{\rm meas}_O\right]$.

\section{Learning Degeneracy on ibm Kyiv}
\label{app:degen}

One possible source of model violation is the choice of how to assign the specific value to Pauli fidelities that can only be learnt as a product in Pauli-learning. We have universally chosen to assume that $f_{P_1} = f_{P_2} = \sqrt{f_{P_1}f_{P_2}}$ for $P_1$ and $P_2$ whose fidelities are degenerate in our learning scheme. We investigate this in Fig.~\ref{fig:kyiv_data_degen}, where we reproduce Fig.~\ref{fig:kyiv_data} but color code in blue those scatter point that correspond to degenerate fidelity pairs. Black points are fidelities that can be learnt uniquely by our Pauli-learning procedure. As can be seen, there is no clear trend towards degenerate pairs having worse systematic error, indicating that this is not the only source of additional model violation beyond the statistical uncertainty ruled out in Fig.~\ref{fig:kyiv_data}.

\section{Further Details of Model Comparison}
\label{sec:cohdrift}

\begin{figure*}[ht]
    \centering
    \includegraphics[width=2\columnwidth]{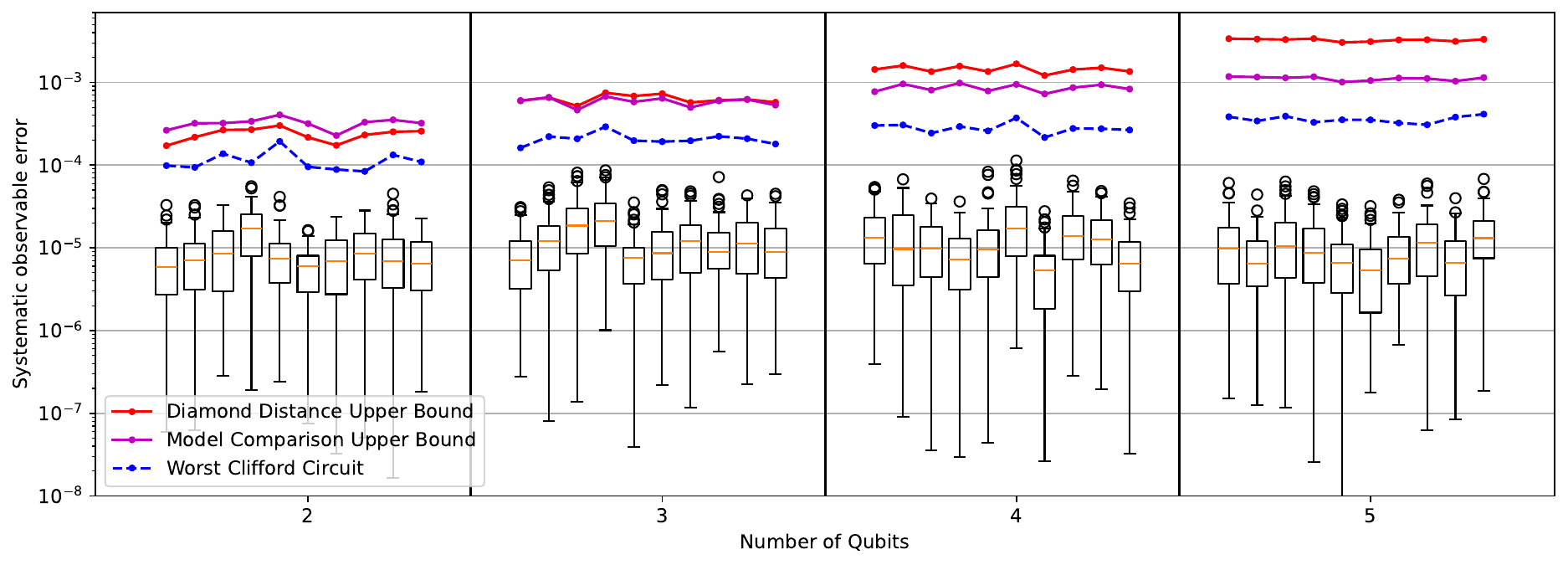}
    \caption{{\bf Learning uncertainty simulations with model comparison.} 
    A reproduction of Fig.~\ref{fig:uncertianty} with the diamond distance upper bound, $\delta_C$, calculated by the model comparison methodology of Eq.~\eqref{eqn:dnorm_comp} shown as a magenta solid line. The red solid line plots the diamond distance upper bound to the systematic error of any circuit, the smaller of $\delta_\gamma$ of Eq.~\eqref{eqn:pecbound_dne} or $\delta_2$ of Eq.~\eqref{eqn:pecbound_2norm}. The blue dashed line plots the worst-case Clifford circuit systematic error of Eq.~\eqref{eqn:pecbound_wce}.}
    \label{fig:uncertainty_sup}
\end{figure*}

We now discuss some further details of using systematic error in PEC to compare error models. Figure \ref{fig:uncertainty_sup} is the same as Fig.~\ref{fig:uncertianty}, but with an added curve plotting the diamond distance upper bound, $\delta_C$, calculated by the model comparison methodology of Eq.~\eqref{eqn:dnorm_comp}. As can be seen, this upper bound calculated by comparing model parameters (generator rates $\lambda$) is typically, but not always, smaller than the upper bounds calculated by comparing the measured and model predicted Pauli fidelities, i.e.~$\delta_\gamma$ of Eq.~\eqref{eqn:pecbound_dne} or $\delta_2$ of Eq.~\eqref{eqn:pecbound_2norm}. Unfortunately, $\delta_C$ can only be calculated when comparing two models, not between a model and the data from which it is derived. The different diamond distance upper bounds can be contrasted in simulation since our simulations have a complete description of the underlying error model.

\subsection{Coherence time drift}

For $n$ qubits with independent $T_1$ values, the mitigation map is parameterized by $2n$ channels $\mathcal{W}_k$, one for Pauli-$X$ type and one for Pauli-$Y$ type error for each qubit. Both of these channels have the same mitigated error rate, which we define as $\Delta_q = \lambda_q -  \hat{\lambda}_q$ for qubit $q$, and as such we will use $\mathcal{W}_{X_q/Y_q}$ to refer to the Pauli-$X/Y$ type single-Pauli components of the mitigated map. The full mitigated map of a single layer is therefore given by
\begin{align}
    \mathcal{V} =  \prod_q \mathcal{W}_{X_q}\mathcal{W}_{Y_q}.
\end{align}
The worst-case Clifford circuit propagates the full weight Pauli-$Z$ operator through each layer, and since the mitigation map factorizes, so too does the Pauli fidelity ratio of $Z^{\otimes n}$
\begin{align}
    r_{Z^{\otimes n}} = \prod_q^n r_{Z_q},
\end{align}
where $r_{Z_q} = \exp\left(-4\Delta_q\right)$ is the Pauli fidelity ratio of the weight-one Pauli with $Z$ on qubit $q$. The elements of Eq.~\eqref{eqn:dnorm_comp} can be a expressed in terms of $r_{Z_q}$ as
\begin{align}
    &\dnorm{\mathcal{W}_{X_q/Y_q}} = \frac{1}{2}\left( \abs{1 + e^{-2\Delta_q}} +  \abs{1 - e^{-2\Delta_q}}\right) \\ 
    & \nonumber= \frac{1}{2}\left( \abs{1 + \sqrt{r_{Z_q}}} +  \abs{1 - \sqrt{r_{Z_q}}}\right) = \max\left(\sqrt{r_{Z_q}},1\right), \\
    &\dnorm{\mathcal{V}} = \prod_q^n \dnorm{\mathcal{W}_{X_q}}\dnorm{\mathcal{W}_{Y_q}} = \prod_q^n\max\left(r_{Z_q},1\right),\\
    &\dnorm{\mathcal{I}-\mathcal{W}_{X_q/Y_q}} = \abs{1 - e^{-2\Delta_q}} = \abs{1-\sqrt{r_{Z_q}}}.
\end{align}

In order to calculate $\dnorm{\mathcal{I} - \mathcal{V}}$ analytically we assume from here on that $r_{Z_q} \geq 1$ for all qubits, as this is the situation where the worst-case Clifford circuit systematic error matches $\delta_C$. With this in mind we can derive that
\begin{align}
    \dnorm{\mathcal{I} - \mathcal{V}} = \sum_j^{2n}(x_j-1)\prod_m^{j-1} x_m,
\end{align}
where the vector 
\begin{align}
    \vec{x} = \left[\sqrt{r_{Z_1}},\sqrt{r_{Z_1}},\sqrt{r_{Z_2}},\sqrt{r_{Z_2}},...,\sqrt{r_{Z_n}},\sqrt{r_{Z_n}}\right]
\end{align}
lists each $\sqrt{r_{Z_q}}$ twice to account for both $\mathcal{W}_{X_q}$ and $\mathcal{W}_{Y_q}$. As before, we use the convention that $\prod_1^0 = 1$. Expanding the expression, it becomes clear that the diamond distance takes on a very simple form
\begin{align}
    \nonumber\dnorm{\mathcal{I} - \mathcal{V}} &= \sum_j^{2n}\prod_m^{j} x_m - \sum_j^{2n}\prod_m^{j-1} x_m = \prod_m^{2n} x_m - 1 \\ 
    &= \prod_m^{n} r_{Z_q} - 1.
\end{align}
The full expression in Eq.~\eqref{eqn:dnorm_comp} for $\delta_C$ follows a similar pattern as above, and we finally arrive at
\begin{align}
    \delta_C = \prod_m^{n} r^l_{Z_q} - 1 = r^l_{Z^{\otimes n}} -  1,
\end{align}
for a circuit with $l$ repetitions of the same $T_1$ mitigated map on $n$ qubits. This is exactly the value of the systematic error for the worst-case Clifford circuit previously discussed, and we see that the worst-case Clifford circuit saturates the diamond distance upper bound for over-corrected mitigation.

\section{Model Violation in Probabilistic Error Amplification}
\label{sec:PEA}

Zero Noise Extrapolation (ZNE) \cite{Temme17,Li17,Endo18} is an error mitigation technique that measures observable expectation values at several points of ``stretched'' or amplified error, such that an extrapolation can be done by numerical curve fitting to the zero-error limit. Probabilistic Error Amplification (PEA) \cite{utility} is a version of ZNE that works for Pauli-stochastic channels, where the error can be stretched by randomly inserting Pauli gates drawn from the correct distributions, and averaging over the resulting circuit randomizations. The sampling distributions are given by the coefficients of the learnt error model for each layer, such that for amplification depth $\mu \in \mathbb{R}^+$ the error at each layer is $\mathcal{E}_j\hat{\mathcal{E}}_j^\mu$. PEA and PEC can be brought under a unified framework if we extend $\mu \in \mathbb{R}$, in which case PEC corresponds to the case $\mu = -1$.

To understand the systematic error in PEA from model violation, we begin by calculating the systematic error at each amplification depth, given by
\begin{align}
    \nonumber\eta(\mu) &= \abs{{\rm Tr}\left[\prod_{j=1}\mathcal{U}_j\mathcal{E}_j\mathcal{E}_j^\mu(\rho)O\right] - {\rm Tr}\left[\prod_{j=1}\mathcal{U}_j\mathcal{E}_j\hat{\mathcal{E}}_j^\mu(\rho)O\right]} \\ 
    &\leq \dnorm{\prod_{j=1}\mathcal{U}_j\mathcal{E}_j\mathcal{E}_j^\mu - \prod_{j=1}\mathcal{U}_j\mathcal{E}_j\hat{\mathcal{E}}_j^\mu}.
\end{align}
Again using the sub-multiplicativity of the diamond norm, we find
\begin{align}
    \eta(\mu) \leq \sum_j \dnorm{\mathcal{E}_j^\mu - \hat{\mathcal{E}}_j^\mu}, 
\end{align}
which has a similar structure to the bound on $\delta_o$ obtained for PEC. However, it scales more favorably as a function of circuit depth, as it does not have the product of $\dnorm{\mathcal{V}_k}$ terms the PEC bound does since all channels in the expression are CPTP. 

To extract an estimate for the upper-bound of systematic error in a mitigated observable, $\delta_o$, we perform the PEA numerical fit with an uncertainty covariance matrix, $\Sigma$, supplied to the fitting algorithm that is built from $\eta(\mu)$ at each amplification depth. $\Sigma$ is given by 
\begin{align}
    \Sigma_{kk} = \eta(\mu_k)^2,~~\Sigma_{jk} = \eta(\mu_j)\eta(\mu_k).
\end{align}
The uncertainty in the output of the fit will thus include the systematic error due to model violation. This covariance matrix assumes that the model violation error is perfectly correlated for each value of $\mu$, or more concretely that $\eta(\mu)$ is a linear function of $\mu$. While generally not the case, this approximation is accurate in the low error regime, and is analogous to the assumptions made to justify a linear extrapolation in PEA/ZNE.

What remains is to upper bound $\eta(\nu)$, and to do so we use the approach of section \ref{subsec:2norm} to upper bound each $\dnorm{\mathcal{E}_j^\mu - \hat{\mathcal{E}}_j^\mu}$ via (suppressing the subscript $j$ for brevity)
\begin{align}
    &\dnorm{\mathcal{E}^\mu - \hat{\mathcal{E}}^\mu} \leq \abs{\nu_0-\hat{\nu}_0} \\&\nonumber+ \frac{4^n-1}{4^n}\sqrt{\sum_{k}\Delta_k\left(\Delta_k - \frac{1}{4^n-1}\sum_{m \neq k}\Delta_m\right)},
\end{align}
where $\Delta_k = f_k - \hat{f}_k$ is the difference between the $k$'th Pauli fidelity $f_k$ of $\mathcal{E}^\mu$ and $\hat{f}_k$ of $\hat{\mathcal{E}}^\mu$, and $\nu_0$ ($\hat{\nu}_0$) is the process fidelity of $\mathcal{E}^\mu$ ($\hat{\mathcal{E}}^\mu$). As was the case for PEC, we can also construct approximate upper bounds for PEA using the worst-case Clifford circuit construction.

\section{Model Violation in Tensor-Network Error Mitigation}
\label{sec:TEM}

The tensor-network error mitigation (TEM) protocol of Ref.~\cite{Filippov23} produces a tensor-network representation of a noisy circuit using the learnt error models of the hard-layers. The layer-by-layer inverse of the noisy circuit combined with the ideal circuit (both in tensor network-form) can then be applied in classical post-processing to the experimental results of the noisy circuit measured in an informationally complete POVM \cite{Perez21}. This enables estimation of mitigated observables, and, due to a middle-out contraction trick in the classical computation of the combined noisy inverse and ideal circuit, the classical post-processing can be efficient if the tensor-network representations of the inverse error channels have low bond dimension. 

Using the notation of this work (but changing to the convention of Ref.~\cite{Filippov23} where the error channel happens after-the-gate), the TEM-mitigated observable expectation value can be expressed as
\begin{align}
    \tilde{o} = {\rm Tr}\left[O \mathcal{C}\prod_{j=l}^1\mathcal{U}^\dagger_j\hat{\mathcal{E}}^{-1}_j\prod_{k=1}^l\mathcal{E}_k\mathcal{U}_k(\rho)\right],
\end{align}
where the ideal circuit $\mathcal{C} = \prod_k^l \mathcal{U}_k$, and we have used a product symbol with a larger lower index than upper index to represent the layer-by-layer noisy inverse circuit. As with PEC, the systematic error is upper bounded by the diamond distance between the ideal and mitigated maps (note that as in PEC the mitigated map need not be CPTP)
\begin{align}
    \delta_o \leq \dnorm{\mathcal{C} - \mathcal{C}\prod_{j=l}^1\mathcal{U}^\dagger_j\hat{\mathcal{E}}^{-1}_j\prod_{k=1}^l\mathcal{E}_k\mathcal{U}_k}.
\end{align}
Following a similar procedure to Appendix \ref{app:DNbounds}, we manipulate this expression as follows to obtain a result analogous to Eq.~\eqref{eqn:PECgen}.
\begin{widetext}
\begin{align}
    \nonumber \delta_o &\leq \dnorm{\mathcal{I} - \prod_{j=l}^1\mathcal{U}^\dagger_j\hat{\mathcal{E}}^{-1}_j\prod_{k=1}^l\mathcal{E}_k\mathcal{U}_k}
    = \dnorm{\mathcal{U}^\dagger_1\mathcal{U}_1 - \mathcal{U}^\dagger_1\hat{\mathcal{E}}^{-1}_1\left(\prod_{j=l}^2\mathcal{U}^\dagger_j\hat{\mathcal{E}}^{-1}_j\prod_{k=2}^l\mathcal{E}_k\mathcal{U}_k\right) \mathcal{E}_1\mathcal{U}_1} \\
    \nonumber &= \dnorm{\mathcal{I} - \hat{\mathcal{E}}^{-1}_1\left(\prod_{j=l}^2\mathcal{U}^\dagger_j\hat{\mathcal{E}}^{-1}_j\prod_{k=2}^l\mathcal{E}_k\mathcal{U}_k\right) \mathcal{E}_1} = \dnorm{\hat{\mathcal{E}}^{-1}_1\hat{\mathcal{E}}_1 - \hat{\mathcal{E}}^{-1}_1\left(\prod_{j=l}^2\mathcal{U}^\dagger_j\hat{\mathcal{E}}^{-1}_j\prod_{k=2}^l\mathcal{E}_k\mathcal{U}_k\right) \mathcal{E}_1} \\
    \nonumber &\leq \dnorm{\hat{\mathcal{E}}^{-1}_1 - \hat{\mathcal{E}}^{-1}_1\left(\prod_{j=l}^2\mathcal{U}^\dagger_j\hat{\mathcal{E}}^{-1}_j\prod_{k=2}^l\mathcal{E}_k\mathcal{U}_k\right)} + \dnorm{\hat{\mathcal{E}}_1 - \mathcal{E}_1}\dnorm{\hat{\mathcal{E}}^{-1}_1} \\
    \nonumber &\leq \left(\dnorm{\mathcal{I} - \left(\prod_{j=l}^2\mathcal{U}^\dagger_j\hat{\mathcal{E}}^{-1}_j\prod_{k=2}^l\mathcal{E}_k\mathcal{U}_k\right)} + \dnorm{\hat{\mathcal{E}}_1 - \mathcal{E}_1}\right)\dnorm{\hat{\mathcal{E}}^{-1}_1} \\
    \nonumber& \leq  \dnorm{\mathcal{I} - \hat{\mathcal{E}}^{-1}_{l}\mathcal{E}_l}\prod_{j=1}^{l-1}\dnorm{\hat{\mathcal{E}}^{-1}_{j}} + \sum_{k=1}^{l-1} \dnorm{\hat{\mathcal{E}}_k - \mathcal{E}_k}\prod_{j=1}^{k}\dnorm{\hat{\mathcal{E}}^{-1}_{j}} \\
    & \leq \dnorm{\mathcal{I} - \hat{\mathcal{E}}^{-1}_{l}\mathcal{E}_l}\prod_{j=1}^{l-1}\dnorm{\hat{\mathcal{E}}^{-1}_{j}} + \sum_{k=1}^{l-1} \dnorm{\mathcal{I} - \hat{\mathcal{E}}^{-1}_{k}\mathcal{E}_k}\prod_{j=1}^{k}\dnorm{\hat{\mathcal{E}}^{-1}_{j}}.
\end{align}
\end{widetext}
As for PEC, this expression depends on a set of diamond distances and norms for which we have developed estimation techniques in the main text. The diamond norm weighting factor per layer (due to the potentially non-CPTP mitigation map) is $\dnorm{\hat{\mathcal{E}}^{-1}_{j}}$, which is strictly greater than or equal to that for PEC, $\dnorm{\mathcal{E}_j\hat{\mathcal{E}}^{-1}_{j}}$. Note that the final layer $l$ only picks up weighting factors for layers up to $l-1$.

\bibliography{OOM_bib}

\end{document}